\documentclass[11pt,onecolumn,draftcls]{IEEEtran} %

\usepackage[linesnumbered, algoruled, longend, boxed]{algorithm2e}
\usepackage{graphicx}
\usepackage{subfigure}
\usepackage{amsmath}
\usepackage{float}
\usepackage{amssymb}
\usepackage{bm}
\usepackage{amsthm}
\usepackage{color}
\usepackage{cite}
\usepackage{url}
\usepackage{setspace}
\usepackage{verbatim}

\newtheorem{theorem}{Theorem}

\newtheorem{lemma}{Lemma}

\newcommand{\stack}[3]{\mathrel{\mathop{\kern0pt #1}\limits_{#3}^{#2}}}

\graphicspath{{fig/}}

\newcommand{\diag}{\textrm{diag}}
\newcommand{\RMS}{\textrm{RMS}}
\newcommand{\bI}{{\mathbf I}}
\newcommand{\bH}{{\mathbf H}}
\newcommand{\bF}{{\mathbf F}}

\newcommand{\bP}{{\mathbf P}}
\newcommand{\bQ}{{\mathbf Q}}
\newcommand{\bV}{{\mathbf V}}
\newcommand{\bR}{{\mathbf R}}

\newcommand{\bX}{{\mathbf X}}

\newcommand{\bx}{{\mathbf x}}
\newcommand{\by}{{\mathbf y}}

\newcommand{\bOmega}{{\mathbf \Omega}}

\def\bxi{{\bm \xi}}
\def\bomega{{\bm \omega}}

\def\bmu{{\bm \mu}}
\def\bR{{\mathbf R}}

\newcommand{\prior}{\mathrm{prior}}
\newcommand{\post}{\mathrm{post}}

\newcommand{\bA}{{\mathbf A}}
\newcommand{\bB}{{\mathbf B}}
\newcommand{\bC}{{\mathbf C}}
\newcommand{\bE}{{\mathbb E}}

\newcommand{\1}{{\textbf 1}}

\newcommand{\ave}{\mathrm{ave}}
\newcommand{\var}{\mathrm{var}}
\newcommand{\cov}{\mathrm{cov}}

\DeclareMathOperator{\tr}{tr}

\begin{document}

\title{Efficient delay-tolerant particle filtering}

\author{{Boris N. Oreshkin, Xuan Liu and Mark J. Coates} \\
Telecommunications and Signal Processing--Computer Networks
Laboratory\\
Department of Electrical and Computer Engineering \\
McGill University, Montreal, QC, Canada \\
Email: {\{boris.oreshkin, xuan.liu2\}@mail.mcgill.ca;
mark.coates@mcgill.ca} }

\maketitle
\vspace{-1cm}
\begin{abstract}

  This paper proposes a novel framework for delay-tolerant particle
  filtering that is computationally efficient and has limited memory
  requirements. Within this framework the informativeness of a delayed
  (out-of-sequence) measurement (OOSM) is estimated using a
  lightweight procedure and uninformative measurements are immediately
  discarded. The framework requires the identification of a threshold
  that separates informative from uninformative; this threshold
  selection task is formulated as a constrained optimization problem,
  where the goal is to minimize tracking error whilst controlling the
  computational requirements. We develop an algorithm that provides an
  approximate solution for the optimization problem. Simulation
  experiments provide an example where the proposed framework
  processes less than 40\% of all OOSMs with only a small reduction in
  tracking accuracy.
\end{abstract}


\section{Introduction}
\label{section::Introduction}

\noindent Tracking is frequently performed using multiple sensor
platforms, with measurements being relayed to a central fusion site
over a wireless network. This can lead to some measurements being
delayed through packet losses or processing delays. The fusion centre
is then faced with out-of-sequence measurements (OOSMs). For
some highly non-linear tracking tasks, the particle filter
significantly outperforms the Extended or Unscented Kalman
Filter (EKFs/UKFs). Incorporating delayed measurements into a particle filter in
an efficient manner can be a challenging task. The goal is to retain
tracking accuracy while minimizing storage and computational
requirements.

In this paper, we propose a novel framework for delay-tolerant
particle filtering that is computationally efficient and has limited
memory requirements. To derive the framework we formulate a
constrained optimization problem of selectively processing only the
most informative OOSMs (those that provide the most reduction in
tracking error), where the constraint specifies a maximum allowable
average computational expenditure. We develop an algorithm that
addresses an approximation of this optimization problem. The method
combines a Gaussian approximation of the current particle filter
distribution and a linearization of the dynamics (similar to the EKF)
to derive a procedure for rapidly predicting the anticipated mean
squared error reduction associated with processing each OOSM. We then
derive a threshold for selecting the ``best'' OOSMs while respecting
the average processing cost constraint. Any measurements which are
deemed insufficiently informative are thus immediately discarded.

We report simulation results for an example tracking
scenario where the proposed algorithm processes only 40\% of all
delayed measurements. The algorithm achieves an accuracy that is
almost equivalent to that achievable by re-running the particle filter
each time a delayed measurement is received, but reduces the
computational cost by a factor of almost two.

\subsection{Related Work}
\label{Sec:Related}

\noindent There has been substantial work on the efficient
incorporation of out-of-sequence measurements OOSMs in Kalman
filters~\cite{Hilton1993,Mallick2001, Bar-Shalom2002, Challa2003,
  Bar-Shalom2004a, Zhang2005, Maskell2006, Shen2009a,
  Zhang2010}. Fewer techniques have been proposed for processing
delayed measurements using particle filters. In~\cite{Orton2001},
Orton et al. propose an approach that stores sets of particles for the
last $\ell$ time steps, where $\ell$ is the predetermined maximum
delay. The algorithm samples new particles at the time step of the
delayed measurement and uses these to update the current particle
weights. This method was improved with a Markov chain Monte Carlo
(MCMC) smoothing step to mitigate the potential problem of degeneracy
in~\cite{Orton2005}. When a large number of particles is needed for
accurate tracking, the algorithm has an excessive storage requirement.

Mallick et al. propose an approximate OOSM particle filter based on
retrodiction in \cite{Mallick2002}. When the filter receives an OOSM, it
retrodicts (predicts backwards) the particles to the time step of the
delayed measurement and uses these particles to update the current weights. The
algorithm in \cite{Zhang2010g} also uses retrodiction, but employs the
Gaussian particle filter of~\cite{Kotecha2003}. Retrodiction requires
a {\em backwards information filter}, i.e. a filter that runs
backwards in time. Constructing such a filter is possible for linear
state dynamics, and these are the systems that are studied
in~\cite{Mallick2002,Zhang2010g}. Recent advances in particle
smoothing~\cite{Orguner2008,Fearnhead2010,Briers2010} can be adopted
to extend the applicability of these techniques to non-linear systems.
However, running the backwards information filter remains a
computationally intensive exercise, equivalent to re-running the
particle filter from the time of the delayed measurement.

In \cite{Orguner2008}, Orguner et al. develop strategies
to reduce both the memory requirements and computational complexity of
OOSM particle filters. They propose a ``storage efficient
particle filter'' that only stores statistics (single mean and covariance) of
the particle set, rather than the particles themselves, at previous
time steps. Auxiliary fixed point smoothers are then employed to
determine the likelihood of the delayed measurement conditioned on
each particle in the current set, and this likelihood is used to
update the weight of each particle. The algorithm can only adjust
particle weights, not change particle locations; this can lead to a
particle degeneracy problem if an OOSM is highly informative and should
induce a significant change in the filtering distribution. Orguner et
al. propose a heuristic of ignoring OOSMs that lead to filter
degeneracy, but this is not satisfactory, since the highly informative
OOSMs are often the most important to process.

The algorithm we propose in this paper involves selective processing
of OOSMs. This was first discussed by Orton and Marrs
in~\cite{Orton2001}; they advocated a heuristic approach of discarding
all measurements that are delayed beyond a constant time, with the
constant to be determined through experiment.  More recently,
selective OOSM processing has been considered by Tasoulis et
al. in~\cite{Tasoulis2010} and in our previous work~\cite{Liu2010}.
Tasoulis et al. proposed a number of heuristic metrics to estimate the
utility of delayed measurements and develop threshold-based tests to
discard measurements of low utility. They incorporate these tests into
three Kalman filtering algorithms that are designed to process delayed
measurements. In~\cite{Liu2010} we proposed a threshold based
procedure to discard uninformative delayed measurements, calculating
their informativeness using mutual information and Kullback-Leibler
distance metrics. We applied our approach in the general non-linear
setting, using a combination of the storage-efficient particle filter
proposed in~\cite{Orguner2008} and a re-run particle filter.

The approach proposed by Tasoulis et al. is developed for the Kalman
Filter and it is difficult to extend to more general filtering
problems with non-linearities. The proposed utility metrics are
heuristic and do not truly capture the potential that each delayed
measurement has to improve the tracking performance. The latter issue
is also a failing of our own work in~\cite{Liu2010}; although mutual
information and Kullback-Leibler distance metrics measure the
potential for information gain, they do not directly assess the
potential reduction in estimation error. Perhaps most importantly,
neither~\cite{Tasoulis2010} nor~\cite{Liu2010} identifies a procedure
for threshold selection, despite the fact that the choice of this
threshold can have a major impact on performance and the appropriate
value is a highly application-sensitive quantity.

\subsection{Paper Organization}
\label{subsection::Paper Organization}

\noindent The rest of the paper is organized as follows.
Section~\ref{Sec:PS} provides a formal problem statement.
Section~\ref{Sec:3} describes memory efficient OOSM particle
filters. Section~\ref{Sec:4} presents the proposed novel framework for
selecting informative OOSMs. In Section~\ref{Sec:5} we explore the
approximations made in the derivation of the framework and present a
theorem identifying asymptotic conditions under which one of the key
approximations becomes exact. Section~\ref{Sec:6} presents a concrete
OOSM particle filtering algorithm based on the selection framework and
Section~\ref{Sec:7} describes simulation experiments for an example
tracking scenario. We make concluding remarks in Section~\ref{Sec:8}.

\section{Problem Statement}\label{Sec:PS}

\noindent We now provide a formal statement of the OOSM filtering
problem that we address and formulate the optimization task. We
consider the general discrete-time Markov state-space model with state
dynamics and measurement models both defined by non-linear maps. The
innovation and observation noises are modelled as additive Gaussian.
At each timestep $k$, there is an active set of distributed sensors,
$\mathcal{V}_k$, that make measurements and $K = \sup_{k\geq1}
|\mathcal{V}_k|$ is the maximal number of active sensors. These
measurements are relayed to the fusion centre. A subset of them
$\mathcal{S}_k$ experience minimal delay and can be processed at time
$k$. Other measurements are delayed and only become available for
processing at later timesteps. Measurements delayed by more than
$\ell$ timesteps are discarded.

The system is described by the following state-space model:
\begin{align} \label{eqn:Markov_model_gen}
X_{k} &= \mathit{f}_{k}(X_{k-1}) + \vartheta_{k}\\
Y_k^s &= \mathit{h}_k^s (X_k)+ \zeta_k^s  \quad (\forall s\in \mathcal{V}_k )\\
\mathcal{Y}_k &= \{ Y_{k}^{\mathcal{S}_{k}} :  \ \mathcal{S}_{k} \subseteq \mathcal{V}_{k} \} \\
\mathcal{Z}_k &= \{ Y_{k-\ell}^{\mathcal{S}_{k-\ell,k}},
Y_{k-\ell+1}^{\mathcal{S}_{k-\ell+1,k}}, \ldots,
Y_{k-1}^{\mathcal{S}_{k-1,k}} \}
\end{align}
Here $\{X_k\}$ denotes the state sequence, which is a Markov diffusion
process with initial distribution $X_0 \sim p(x_0)$, and $\{Y_k^s\}$
denotes the measurement sequence at the $s$-th sensor, with
$Y_{k}^{\mathcal{S}_{k}} = \{ Y_k^s : s \in \mathcal{S}_{k} \}$.
$\vartheta_{k}$ is the innovation noise with Gaussian distribution
$\mathcal{N}(0,\bV_{k})$, and $\zeta_k^s$ is the measurement noise
with Gaussian distribution $\mathcal{N}(0,\bQ^s_k)$. The functions
$\mathit{f}_{k} : \mathbb{R}^d \rightarrow \mathbb{R}^d$ and
$\mathit{h}^s_k : \mathbb{R}^{d} \rightarrow \mathbb{R}^{m_s}$ are the
state transition and measurement maps. $\mathcal{Y}_k$ denotes the set
of non-delayed measurements received at time $k$.  $\mathcal{Z}_k$
denotes the set of OOSMs received at time $k$. The set
$\mathcal{S}_{\tau,k}$ is the subset of active sensors at time $\tau$
whose measurements are received at time step $k$ ($\mathcal{S}_{k,k}
\equiv \mathcal{S}_{k}$); $Y_{\tau}^{\mathcal{S}_{\tau,k}}$ is the set
of measurements made at time $\tau$ that arrive at the fusion centre
at time $k$.

\subsubsection{OOSM Filtering}

Let $\mathcal{W}_{i:j, k}$ denote the set of measurements generated in
the interval $[i,j]$ available at the fusion centre by time $k$.  This
includes all the non-delayed measurements $\mathcal{Y}_{i:j} =
\bigcup_{m=i}^j \mathcal{Y}_{m}$ and OOSMs $\mathcal{Z}_{i:j, k} =
\{Z_{\tau, m}^{s} \in \mathcal{Z}_{m} : \tau \in [i,j], s \in
\mathcal{S}_{\tau,m}, m \in [i+1, k] \}$, where $Z_{\tau, m}^{s}$ is
the OOSM that was acquired at time $\tau$ by the sensor $s \in
\mathcal{S}_{\tau,m} \subseteq \mathcal{V}_{\tau}$ and was received at
the fusion centre at time $m$. Let $\widetilde{\mathcal{W}}_{i:j,k} =
\mathcal{W}_{i:j,k} \setminus\mathcal{Z}_{k} =
\{\mathcal{W}_{i:j,k-1},\mathcal{Y}_k\}$, i.e. the set of all
measurements available at time $k$ except those in
$\mathcal{Z}_{k}$. Lastly, note that $\mathcal{Z}_{\tau, k} \equiv
\mathcal{Z}_{\tau:\tau, k}$, $\mathcal{W}_{\tau,k} \equiv
\mathcal{W}_{\tau:\tau,k}$ and $\widetilde{\mathcal{W}}_{\tau,k}
\equiv \widetilde{\mathcal{W}}_{\tau:\tau,k}$.

The sequential OOSM filtering task involves calculating an estimate
$\widehat{X}_k$ of the current state, given all available measurements
at time $k$, $\mathcal{W}_{1:k,k}$. In this work, we form the estimate
by calculating an approximate expectation of the state by sequentially
calculating a particle representation of the posterior distribution.

\subsubsection{Selective Processing for Computational Constraints}

In this paper we are interested in reducing computational requirements
by processing only the informative OOSMs. We formulate this problem as
an optimization problem that involves minimizing the mean-squared
error (with respect to an $L_2$ norm) subject to satisfying a
constraint ($C_{ave}$) on the expected computation at each time step.

Let $b_{\tau,k}^{s} \in \{ 0,1 \}$ be the indicator of OOSM
$Z_{\tau,k}^{s}$ arrival and denote by $p_{\tau,k}^{s}$ the expected
value of $b_{\tau,k}^{s}$, conditioned on all the measurements
received prior to time $k$. Denote by $C_{\tau,k}^{s}$ the
computational cost associated with processing the OOSM
$Z_{\tau,k}^{s}$.  Let $d_{\tau,k}^{s} \in \{ 0,1 \}$ be our decision
to process or reject measurement $Z_{\tau,k}^{s}$ and $\mathcal{D}_k =
\bigcup_{\tau=k-\ell}^{k-1} \bigcup_{s \in \mathcal{S}_{\tau,k}} \{
d_{\tau,k}^{s} \}$ be the current set of all possible decisions.
Decisions must be made sequentially, prior to the arrivals of the
OOSMs at time $k$ due to the real-time nature of the tracking
task. The goal is to ensure that the computational constraint is
obeyed {\em on average} at each time step, i.e. in expectation with respect
to all possible arrivals of OOSMs.

We thus address the following optimization task for each $k$ over the
tracking period:
\begin{align}
&\min_{\mathcal{D}_k \in \{0,1\}^{\sum_{\tau}|\mathcal{V}_{\tau}|}}
\bE\left\{|X_{k}-\widehat X_{k}|^2\right\}  \nonumber
\\ &\text{subject to} \quad
\sum_{k=1}^K\sum_{\tau=k-\ell}^{k-1} \sum_{s\in \mathcal{D}_{\tau,k}}
d_{\tau,k}^{s} p_{\tau,k}^{s} C_{\tau,k}^{s} \leq KC_{\ave}
\end{align}

\section{OOSM Particle Filters} \label{Sec:3}

\noindent Previously proposed OOSM particle filters primarily differ
in how they incorporate the OOSMs from the set $\mathcal{Z}_k$. The
simplest approach is to discard them, but this often results in poor
tracking performance. Another obvious approach is to restart the
filter at the time step immediately prior to the time step associated
with the earliest OOSM in $\mathcal{Z}_k$ and re-run to the current
time step $k$. This requires that we record all the particles, weights
and the measurements for the maximal delay window. We call this
approach the ``OOSM re-run particle filter'' and consider it to be an
accuracy benchmark. This method has two unattractive qualities: the
storage requirements can be immense and the computation cost is high.

As discussed in Section~\ref{Sec:Related}, several methods have been
proposed to alleviate these costs. In this section, we provide a brief
review of the storage efficient particle filter of~\cite{Orguner2008}
and describe a relatively obvious alternative algorithm that we
introduced in~\cite{Liu2010}. In both algorithms, the memory
requirements are reduced by storing statistics of the particle sets
from past time steps instead of the particles themselves. The past
particle distributions are approximated by Gaussian approximations. The
stored information is then the mean and covariance matrix of particles
at each time step from $k-\ell-1$ to $k$. Denote, respectively, by
$\bxi_{k}$, $\bomega_{k}$ the sets of the values and weights of
particles at time $k$, and let $\bmu_{k}$, $\bR_{k}$ denote their mean
and covariance. The stored information is then
\begin{align}
\bOmega_k &= \{  \bmu_{k-\ell:k}, \bR_{k-\ell:k},
\mathcal{W}_{k-\ell-1:k,k} \}, \label{E:Omega}
\end{align}
Here $\bmu_{k-\ell-1:k}$ and $\bR_{k-\ell-1:k}$ denote, respectively,
the means and covariances of the particle sets for time-steps ranging from
$k-\ell-1$ to $k$.

A generic storage efficient OOSM particle filtering algorithm is
summarized in Algorithm~\ref{alg:PFOOSM}. If there are no OOSMs at
time $k$, we write $\mathcal{Z}_k = \emptyset$.

\begin{algorithm}[H] \label{alg:PFOOSM}

\SetKwFunction{ProcessOOSM}{ProcessOOSM}
\SetKwFunction{ParticleFilter}{ParticleFilter}
\SetKwFunction{SaveGauss}{SaveGauss}

At time $k$

\KwIn{ $\mathcal{Z}_k$, $\bOmega_{k-1}$, $\bxi_{k-1}$, $\bomega_{k-1}$,$\mathcal{Y}_k$ } 

($\bxi_{k}$, $\bomega_{k}$) $\leftarrow$
        \ParticleFilter{$\mathcal{Y}_k$, $\bxi_{k-1}$, $\bomega_{k-1}$} \;

($\bmu_{k}$, $\bR_{k}$) $\leftarrow$
        \SaveGauss{$\bxi_{k}$, $\bomega_{k}$} \;
\uIf{$\mathcal{Z}_k \neq \emptyset$}{

    ($\bxi_{k}$, $\bomega_{k}$, $\bOmega_k$) $\leftarrow$
        \ProcessOOSM{$\mathcal{Z}_k$, $\bxi_{k}$, $\bomega_{k}$, $\bOmega_k$}\; 

}

\caption{Generic OOSM Particle Filter}
\end{algorithm}
In this algorithm, the function \verb,ParticleFilter, can be any
standard particle filtering method. If $\mathcal{Y}_k=\emptyset$,
\verb,ParticleFilter, only propagates the particles and skips the
measurement processing step. The function \verb,SaveGauss, calculates the
maximum likelihood estimates of the mean and covariance given the
weighted sample set $\bxi_{k}$, $\bomega_{k}$ and stores these in $\bOmega_k$:
\begin{align}
\bmu_{k} &= \sum_{i=1}^N \bomega_{k}^{(i)} \bxi_{k}^{(i)}
\\
\bR_{k} &= \sum_{i=1}^N \bomega_{k}^{(i)} (\bxi_{k}^{(i)}-\bmu_{k})
(\bxi_{k}^{(i)}-\bmu_{k})^T
\end{align}
The function \verb,ProcessOOSM, specifies how OOSMs are processed and
varies depending on the specific algorithm.

\subsection{Gaussian Approximation Re-run Particle Filter
(OOSM-GARP)} \label{Sec:OOSM-GARP}

\noindent A simple modification of the re-run particle filter
involves storing only Gaussian approximations of the particle
distributions at previous timesteps. When a batch of OOSMs arrives,
the particle filter is re-run from the time step preceding the
earliest OOSM. Since the particle set from that time step is
unavailable, particles are generated from the stored approximation.

When OOSM-GARP receives $\mathcal{Z}_k$ at time $k$, it returns to
the time step $\widetilde\tau_k-1$ (let $\widetilde\tau_k$ denote
the earliest time step of all OOSMs in $\mathcal{Z}_k$). It samples
particles from $\mathcal{N}(\bmu_{\widetilde\tau_k-1},
\bR_{\widetilde\tau_k-1})$, propagates them to the time step
$\widetilde\tau_k$ and runs the filter as standard particle filter
using all stored measurements. At each step, it
updates the mean and covariance matrix in the stored set $\bOmega_k$
as described in Algorithm~\ref{alg:ProcessOOSM-GARP}.

\begin{algorithm}[H] \label{alg:ProcessOOSM-GARP}

\SetKwFunction{SaveGauss}{SaveGauss}
\SetKwFunction{ParticleFilter}{ParticleFilter}
\SetKwFunction{ProcessMeasurement}{ProcessMeasurement}

\KwIn{  $\mathcal{Z}_k$, $\bOmega_k$ }

$\widetilde\tau_k = \min\limits_{\tau}  \{ \tau : y_\tau \in
\mathcal{Z}_k \}$ \;

$\{\bxi_{\widetilde\tau_k-1}^{(i)}\}_{i=1}^N$ $\sim$
$\mathcal{N}(x_{\widetilde\tau_k-1}, \bmu_{\widetilde\tau_k-1},
\bR_{\widetilde\tau_k-1})$ \;

$\bomega_{\widetilde\tau_k-1}^{(i)} = 1/N, \ i = 1\ldots N$ \;

\For{$j = \widetilde\tau_k, \ldots, k $}{

    ($\bxi_{j}$, $\bomega_{j}$) $\leftarrow$
        \ParticleFilter{$\mathcal{W}_{j,k}$, $\bxi_{j-1}$, $\bomega_{j-1}$}\;
    ($\bmu_{j}$, $\bR_{j}$) $\leftarrow$
        \SaveGauss{$\bxi_{j}$, $\bomega_{j}$} \;

} \caption{ ProcessOOSM-GARP }
\end{algorithm}

In many tracking tasks, the Gaussian provides a reasonable
approximation to the particle distributions. In OOSM-GARP, the
Gaussian is only used to re-start the particle filter (to draw initial
samples), so the impact of approximation errors on filtering
performance is relatively small. OOSM-GARP thus performs
almost as well as the basic re-run particle filter but requires much
less memory. However, OOSM-GARP is relatively computationally
complex since it reprocesses all the particles for
$k-\widetilde\tau_k+1$ steps. Note that the cost to process OOSMs
corresponding to a single time step, $\mathcal{Z}_{j, k}$, is approximately equal to that
of processing the whole batch of OOSMs $\mathcal{Z}_{j:k-1, k}$
since we have to execute the particle filter from time $j$ to time $k$ in
both cases. This cost is proportional to the total computational
complexity of functions \verb,ParticleFilter, and \verb,SaveGauss,
multiplied by a factor of $k-j$.

\subsection{Storage Efficient Particle Filter with EKS (SEPF-EKS)} \label{Sec:SEPFEKS}

\noindent We now provide a brief review of the storage efficient OOSM
particle filter from~\cite{Orguner2008}. Orguner et al. described
three versions of the filter, which differed according to the
auxiliary fixed-point smoother they employed. We focus on the filter
that employs Extended Kalman Smoother, since it is the least
computationally demanding but has comparable tracking performance.

The SEPF is based on the following weight-update equation:
\begin{align}
\bomega_{k}^{(i)} &\propto
p(\mathcal{Z}_{\tau,k}|\bxi_k^{(i)},\widetilde{\mathcal{W}}_{{1:k},k})
\bomega_{k,\bar{\tau}}^{(i)} \label{E:wup} .
\end{align}
Here $\bomega_{k,\bar{\tau}}^{(i)}$ and $\bomega_{k}^{(i)}$ denote
the weights before and after processing $\mathcal{Z}_{\tau,k}$.
The SEPF estimates this likelihood expression in two stages. First
it approximates $p(x_{\tau} | \bxi_k^{(i)},
\widetilde{\mathcal{W}}_{{1:k},k})$ by applying an
augmented-state extended Kalman smoother~\cite{Biswas1973}, treating
the current particle $\bxi_k^{(i)}$ as a measurement. The SEPF then
employs an EKF approximation of $p(\mathcal{Z}_{\tau,k}|x_{\tau})$
to construct an estimate of the likelihood $p(\mathcal{Z}_{\tau,k} |
\bxi_k^{(i)}, \widetilde{\mathcal{W}}_{{1:k},k})$.

Although the original algorithm was designed to treat individual
OOSMs, it can be easily extended to treat batches of OOSMs by
running a separate update for each time-step. This extended
algorithm is presented as Algorithm~\ref{alg:SEPF-EKS}.

\begin{algorithm}[H] \label{alg:SEPF-EKS}

\SetKwFunction{SaveGauss}{SaveGauss}
\SetKwFunction{ParticleFilter}{ParticleFilter}
\SetKwFunction{ProcessMeasurement}{ProcessMeasurement}

\KwIn{  $\mathcal{Z}_k$, $\bOmega_k$, $\bomega_{k,\bar{\tau}}$, $\bxi_k$}

\For{$\mathcal{Z}_{\tau,k} \in \mathcal{Z}_k$}{

    Compute approximation $p(\mathcal{Z}_{\tau,k} | \bxi_k^{(i)},
    \widetilde{\mathcal{W}}_{{1:k},k})$ for all $i$\;

    $\bomega_{k}^{(i)}$ $\leftarrow$ $\bomega_{k,\bar{\tau}}^{(i)} p(\mathcal{Z}_{\tau,k} | \bxi_k^{(i)}, \widetilde{\mathcal{W}}_{{1:k},k})$ $\forall i$ \;

}

$\bomega_{k} = \bomega_{k} / \sum_{i} \bomega_{k}^{(i)}$ \;

\caption{ ProcessSEPF-EKS }
\end{algorithm}

SEPF-EKS achieves significant computational savings because the
filtering operations for step $2$ are common to all $N$ particles
except for a single time-step. This means that the effective
computational cost is equivalent to running one time step of a
particle filter, and is therefore usually less than that of the
OOSM-GARP filter. The advantage diminishes when it is
common for OOSMs to arrive in batches with different delays because of
the seemingly unavoidable loop in the algorithm.
\section{Selective OOSM Processing} \label{Sec:4}

The computational cost of processing an OOSM is relatively high and
frequently it is wasted effort, resulting in minimal change to the
filtering distribution or the tracking accuracy. In this section we
design a procedure for addressing the optimization problem posed in
Section~\ref{Sec:PS}, that of minimizing the mean squared error while
controlling the computational effort.

The optimization problem is challenging and generating an exact
solution would be more costly than simply processing all OOSMs with a
re-run particle filter. We therefore strive to approximate the problem
so that we can develop an efficient procedure for selecting the
informative OOSMs. The complexity of this procedure must not depend on
the number of particles in the filter.

Our method employs a Gaussian approximation of the joint distribution
of the current state and the current set of OOSMs. We derive this
approximation using an EKF-type linear approximation of the general
state-space model. Second, we model the OOSMs from different sensors
or different times as approximately {\em unconditionally}
independent. This second approximation allows us to disentangle the
effects of processing different OOSMs on the filtering error. In
section~\ref{Sec:5} we study asymptotic conditions under which this assumption
holds exactly. This provides a solid theoretical justification for our
choice of this simplifying approximation and we consider that it is
sufficiently accurate in practice for our purpose of selecting the
informative OOSMs. It is important to stress that these approximations are
only used for the purpose of selecting the measurements to process;
they are not employed within the filter itself.

\subsection{Tracking MSE Under Gaussian Approximation}

\noindent We employ the well known EKF-type linear approximation of
the general state-space model:
\begin{align} \label{eqn:state_space_lin}
X_{k} &= f_k(\mu_{X_{k-1}}) + \bF_{k}(X_{k-1} - \mu_{X_{k-1}}) + \vartheta_{k} \\
Y_{k}^{s} &= h_k^{s}(\mu_{X_{k}}) + \bH_{k}^{s}(X_{k} - \mu_{X_{k}})
+ \zeta_{k}^{s}, \quad \quad \quad s \in \mathcal{V}_k
\end{align}
Here $\bF_{k}$ and $\bH_{k}^{s}$ are linearizations (through Taylor
expansion at $\mu_{X_{k-1}}$ and $\mu_{X_{k}}$, respectively) of the
non-linear dynamic and measurement maps.

Let $\bP_{k}$ and $\mu_{k}$ be the covariance matrix and the mean of
the Gaussian approximation of the joint probability distribution of
the current state and the current set of OOSMs conditioned on all
available measurements.  The covariance matrix and the mean have the
following structure:
\begin{align}
\bP_{k} = \begin{pmatrix} \bR_{X_{k} X_{k}  | \widetilde{\mathcal{W}}_{1:k,k}} & \bR_{X_k \mathcal{Z}_{k}  | \widetilde{\mathcal{W}}_{1:k,k}} \\
\bR_{\mathcal{Z}_{k} X_{k}  | \widetilde{\mathcal{W}}_{1:k,k}} &
\bR_{\mathcal{Z}_{k} \mathcal{Z}_{k}  |
\widetilde{\mathcal{W}}_{1:k,k}}
\end{pmatrix}, \, \mu_{k} = \begin{pmatrix} \mu_{X_{k}| \widetilde{\mathcal{W}}_{1:k,k}} \\ \mu_{\mathcal{Z}_{k}|\widetilde{\mathcal{W}}_{1:k,k}} \end{pmatrix}
\label{eq:covariances}
\end{align}
where $\bR_{X_{k} X_{k} | \widetilde{\mathcal{W}}_{1:k,k}}$ is the
current state covariance, $\bR_{X_{k} \mathcal{Z}_{k} |
  \widetilde{\mathcal{W}}_{1:k,k}}$ and $\bR_{\mathcal{Z}_{k} X_{k} |
  \widetilde{\mathcal{W}}_{1:k,k}} = \bR_{X_{k} \mathcal{Z}_{k} |
  \widetilde{\mathcal{W}}_{1:k,k}}^T$ is the state-measurement
cross-covariance and $\bR_{\mathcal{Z}_{k} \mathcal{Z}_{k} |
  \widetilde{\mathcal{W}}_{1:k,k}}$ is the measurement set covariance.
Note that the means and covariances are conditioned on
$\widetilde{\mathcal{W}}_{1:k,k}$ which includes the current set of
undelayed measurements $\mathcal{Y}_k$ as well as all the OOSMs and
undelayed measurements that have been incorporated up to time $k$. In
the following discussion, we will often skip this conditioning to
avoid unnecessarily complicated notation, but this conditioning is
implied unless explicitly stated otherwise.

The optimal MMSE estimator $\widehat X_{k}$ of the state is known to
be the conditional mean $\mu_{X_{k}|\widetilde{\mathcal{W}}_{1:k,k},
\mathcal{Z}_{k}}$, which in the case of our Gaussian approximation is
simply:
\begin{align}
\widehat X_{k} = \mu_{X_{k}} + \bR_{X_{k} \mathcal{Z}_{k}}
\bR_{\mathcal{Z}_{k} \mathcal{Z}_{k}}^{-1} (\mathcal{Z}_{k} -
\mu_{\mathcal{Z}_{k}}).
\end{align}
Let $\mathcal{B}_k =
\bigcup_{\tau=k-\ell}^{k-1} \bigcup_{s\in\mathcal{S}_{\tau,k}} \{
b_{\tau,k}^{s} \}$ be the set of random variables that indicate OOSM arrivals at time
$k$. This set defines the structure of the set $\mathcal{Z}_{k}$ along with
the associated mean $\mu_{\mathcal{Z}_{k}}$ and (cross-)covariance
terms $\bR_{X_{k} \mathcal{Z}_{k}}$ and $\bR_{\mathcal{Z}_{k}
\mathcal{Z}_{k}}$. By the law of total variance the variance of the
the estimator can be expressed as:
\begin{align}
\var(X_{k}-\widehat X_{k}) = &\bE\{ \var(X_{k}-\widehat X_{k} |
\mathcal{B}_{k}) \} + \var(\bE\{ X_{k}-\widehat X_{k}| \mathcal{B}_{k} \}).
\end{align}
Since, according to our linearization, $\bE\{ X_{k} |
\mathcal{B}_{k} \} = f_k(\mu_{X_{k-1}})$ and $\bE\{ \widehat X_{k} |
\mathcal{B}_{k} \} = \mu_{X_{k}} = f_k(\mu_{X_{k-1}})$ we have for
any realization of $\mathcal{B}_{k}$: $\bE\{ X_{k}-\widehat X_{k}|
\mathcal{B}_{k} \} = 0$. Thus the variance of the MMSE estimator is
equal to the expectation of its variance conditioned on the
realization of indicators $\mathcal{B}_{k}$:
\begin{align}
\var(X_{k}-\widehat X_{k}) = \bE\{ \var(X_{k}-\widehat X_{k} |
\mathcal{B}_{k}) \}.
\end{align}
For a specific realization of indicators $\mathcal{B}_{k}$ this variance is
defined by the components of the joint covariance matrix (recall
that $\mathcal{Z}_{k}$ is a function of $\mathcal{B}_{k}$):
\begin{align}
\var(X_{k}-\widehat X_{k} | \mathcal{B}_{k}) = \bR_{X_{k} X_{k}} -
\bR_{X_{k} \mathcal{Z}_{k}} \bR_{\mathcal{Z}_{k}
\mathcal{Z}_{k}}^{-1} \bR_{\mathcal{Z}_{k} X_{k}}
\end{align}
The mean squared error of estimating the state $X_{k}$ conditioned
on the OOSM set $\mathcal{Z}_{k}$ (as well as all the previous
measurements) is thus given by
\begin{align}
\tr \var(X_{k}-\widehat X_{k}) &= \bE\{\tr
\var(X_{k}-\widehat X_{k} | \mathcal{B}_{k})\}\\
&= \tr\bR_{X_{k} X_{k} } - \bE\{ \tr \bR_{X_{k} \mathcal{Z}_{k}}
\bR_{\mathcal{Z}_{k} \mathcal{Z}_{k}}^{-1} \bR_{\mathcal{Z}_{k}
X_{k}} \}
\end{align}
Under the assumption that the measurements made by different sensors
(or the same sensor at different times) are approximately
unconditionally independent, $\bR_{\mathcal{Z}_{k} \mathcal{Z}_{k}}$ is
approximately block-diagonal. This implies that we can approximate
the above expression as follows:
\begin{align}
&\tr \var(X_{k}-\widehat X_{k}) \approx \tr\bR_{X_{k} X_{k}} 
- \bE\left\{ \sum_{\tau=k-\ell}^{k-1} \sum_{s\in
\mathcal{V}_{\tau}} d_{\tau,k}^{s} b_{\tau,k}^{s} \tr \bR_{X_{k}
Y_{\tau}^{s} } \bR_{Y_{\tau}^{s}Y_{\tau}^{s} }^{-1}
\bR_{Y_{\tau}^{s} X_{k} } \right\}.
\end{align}
Here the expectation is taken with respect to the measurement
arrival indicators $b_{\tau,k}^{s}$, $\bR_{Y_{\tau}^{s}Y_{\tau}^{s}
} = \var( Y_{\tau}^{s}  )$ is measurement covariance and
$\bR_{Y_{\tau}^{s}X_{k} } = \cov( Y_{\tau}^{s}, X_{k}
 )$ is the state-measurement cross
covariance. If we denote
\begin{align} \label{eqn:measurement_utility}
R_{\tau,k}^{s} = \tr \bR_{X_{k} Y_{\tau}^{s} }
\bR_{Y_{\tau}^{s}Y_{\tau}^{s} }^{-1} \bR_{Y_{\tau}^{s} X_{k} },
\end{align}
the factor that we will refer to as the measurement utility then the
expression for the MSE can be further simplified:
\begin{align} \label{eqn:MSE_simplified}
\tr \var(X_{k}-\widehat X_{k}) \approx \tr\bR_{X_{k} X_{k} } - \sum_{\tau=k-\ell}^{k-1} \sum_{s\in
\mathcal{V}_{\tau}} d_{\tau,k}^{s} p_{\tau,k}^{s} R_{\tau,k}^{s}.
\end{align}
where $p_{\tau,k}^{s} = \bE\left\{b_{\tau,k}^{s}\right\}$ is the
probability that the measurement acquired by sensor $s$ at time $\tau$
arrives at time $k$ (conditioned on the measurement arrivals up to
time $k$). The above expression is a natural objective
function to be minimized to assure the best tracking quality. The
minimal value of the objective is reached when all measurements are
processed ($d_{\tau,k}^{s}=1, \forall\, \tau,s$) since $R_{\tau,k}^{s}
\geq 0$.

\subsection{One-step Constrained Minimization of Approximate MSE}

Given the discussion above and the identified approximations, the
constrained optimization problem posed in Section~\ref{Sec:PS} can be formulated
as follows:
\begin{align}
&\min_{\mathcal{D}_k \in \{0,1\}^{\sum_{\tau}|\mathcal{V}_{\tau}|}}
\tr \var(X_{k}-\widehat X_{k}) \nonumber \\
&\quad \text{subject to} \quad
\sum_{\tau=k-\ell}^{k-1} \sum_{s\in \mathcal{D}_{\tau,k}}
d_{\tau,k}^{s} p_{\tau,k}^{s} C_{\tau,k}^{s} \leq C_{\ave} \label{eqn:cost_constraint}
\end{align}
The unconstrained objective to be minimized can be formulated using
Lagrange relaxation with Lagrange multiplier $\gamma_k$:
\begin{align}
J(\mathcal{D}_k) &= \tr \var(X_{k}-\widehat X_{k}) + \gamma_k\left(
\sum_{\tau=k-\ell}^{k-1} \sum_{s\in \mathcal{V}_{\tau}}
d_{\tau,k}^{s} p_{\tau,k}^{s} C_{\tau,k}^{s} - C_{\ave}\right) \nonumber\\
&= \tr\bR_{X_{k} X_{k} } - \sum_{\tau=k-\ell}^{k-1} \sum_{s\in
\mathcal{V}_{\tau}} d_{\tau,k}^{s} p_{\tau,k}^{s} (R_{\tau,k}^{s} -
\gamma_k C_{\tau,k}^{s}) - \gamma_k C_{\ave}.
\end{align}
For a fixed $\gamma_k$ the optimal solution can be found by
optimizing each $d_{\tau,k}^{s}$ independently since the
contribution of each term under the sum corresponding to a
particular $d_{\tau,k}^{s}$ is independent of all other variables to
be optimized. It is clear that setting $d_{\tau,k}^{s} = 1$ whenever
$R_{\tau,k}^{s} - \gamma_k C_{\tau,k}^{s} \geq 0$ and
$d_{\tau,k}^{s} = 0$ whenever $R_{\tau,k}^{s} - \gamma_k
C_{\tau,k}^{s} < 0$ produces the smallest value of the objective
function for a given $\gamma_k$. Substituting this solution into the
constraint we obtain
\begin{align}
\sum_{\tau=k-\ell}^{k-1} \sum_{s\in \mathcal{V}_{\tau}} \1_{\{
R_{\tau,k}^{s} - \gamma_k C_{\tau,k}^{s} \}} p_{\tau,k}^{s}
C_{\tau,k}^{s} \leq C_{\ave},
\end{align}
where $\1_{\{\cdot \}}$ is the indicator function. If we denote
$\widetilde R_{\tau,k}^{s} = R_{\tau,k}^{s} / C_{\tau,k}^{s}$, the
measurement utility diminished by the processing cost incurred, the
above is equivalent to
\begin{align}
\sum_{\{s,\tau :  \widetilde R_{\tau,k}^{s} \geq \gamma_k \}}
p_{\tau,k}^{s} C_{\tau,k}^{s} \leq C_{\ave}. \label{eqn:gamma_ineq}
\end{align}
The optimal value of $\gamma_k$ is thus the smallest value for which
\eqref{eqn:gamma_ineq} holds. A simple practical algorithm can be
devised to identify this value of $\gamma_k$. The algorithm,
summarized in Algorithm~\ref{alg:th_select}, assumes that we can
evaluate $p_{\tau,k}^{s}$, which is usually possible given sufficient
knowledge about the measurement apparatus and the network delay
profile.

\vspace{0.5cm}
\begin{algorithm}[H] \label{alg:th_select}

\KwIn{$\{\widetilde R_{\tau,k}^{s} \}$, $\{ p_{\tau,k}^{s} \}$, $\{
C_{\tau,k}^{s} \}$ of cardinality $T = \sum_{\tau=k-\ell}^{k-1} |
\mathcal{V}_{\tau}|$ and $C_{\ave}$ \;}

Order set $\{\widetilde R_{\tau,k}^{s} \}$ by decreasing value,
output ordered sequence $\{ R_{n}^{o} \}_{n=1}^{T}$ \;

Construct sequences $\{ p_{n}^{o} \}_{n=1}^{T}$, $\{ C_{n}^{o} \}_{n=1}^{T}$
using mapping $(\tau,s) \mapsto n$ used for the previous set \;

Construct sequence $\{ \Psi_{n}^{o} \}_{n=1}^{T}$ with elements
$\Psi_{n}^{o} = \sum_{j=1}^{n} p_{j}^{o} C_{j}^{o}$

Identify $n^* = \arg \max_n \Psi_{n}^{o} : \Psi_{n}^{o} \leq C_{ave}$ 

\KwOut{$\gamma_k = R_{n^*}^{o}$ \;}

\caption{Threshold selection algorithm}
\end{algorithm}

We can now describe the operation of the proposed OOSM selection
algorithm. At every filtering step the selection algorithm first
calculates the measurement utilities diminished by the processing
cost, $\widetilde R_{\tau,k}^{s}$, along with probabilities of arrival
for all possible OOSMs, $p_{\tau,k}^{s}$. It then identifies a
threshold $\gamma_k$ such that the expected processing cost does not
exceed $C_{ave}$ (step 4 in Algorithm~\ref{alg:th_select}). The final
step of the algorithm is to select arriving OOSMs with utility
$\widetilde R_{\tau,k}^{s}$ surpassing the calculated threshold.

To execute the proposed algorithm we need expressions for the (cross-)
covariance matrices $\bR_{X_{k} Y_{\tau}^{s} }$ and
$\bR_{Y_{\tau}^{s}Y_{\tau}^{s} }$.  These matrices can be calculated
online using the extended Kalman smoother (EKS) algorithm. We employ
the Rauch-Tung-Striebel (RTS) backward recursion
realization~\cite{Rauch1965}. We apply the RTS recursion starting from
the Gaussian approximation of the posterior at the current time $k$
and moving backwards in time until time step $k-\ell$. As a result, we
obtain a sequence of smoother means
$\mu_{X_{\tau}|\widetilde{\mathcal{W}}_{1:k,k}}$ and covariance
matrices $\bR_{X_{\tau} X_{\tau}}$ for $k-\ell \leq \tau < k$.

At time $k$ we have the set of measurements
$\widetilde{\mathcal{W}}_{1:k,k}$, so the
linearizations~\eqref{eqn:state_space_lin} can be made more general
(and, hopefully, accurate) with the use of the EKS statistics
$\mu_{X_{\tau}|\widetilde{\mathcal{W}}_{1:k,k}}, \tau<k-1$:
\begin{align} \label{eqn:state_space_lin_EKS}
X_{\tau} &= f_{\tau}(\mu_{X_{\tau-1}|\widetilde{\mathcal{W}}_{1:k,k}}) + \bF_{\tau}(X_{\tau-1} - \mu_{X_{\tau-1}|\widetilde{\mathcal{W}}_{1:k,k}}) + \vartheta_{\tau} \\
Y_{\tau}^{s} &=
h_{\tau}^{s}(\mu_{X_{\tau}|\widetilde{\mathcal{W}}_{1:k,k}}) +
\bH_{\tau}^{s}(X_{\tau} -
\mu_{X_{\tau}|\widetilde{\mathcal{W}}_{1:k,k}}) + \zeta_{\tau}^{s},
\ s \in \mathcal{V}_{\tau}.
\end{align}
Here the Jacobians $\bF_{\tau}$ and
$\bH_{\tau}^{s}$ are evaluated at the points defined by
the respective EKS means. With the use of the above linearization,
calculation of the required approximate covariance matrices becomes
straightforward. Noting that $\bE\{ Y_{\tau}^{s} \} =
h_{\tau}^{s}(\mu_{X_{\tau}})$, and observing the independence of
$\zeta_{\tau}^{s}$ and $X_{\tau} - \mu_{X_{\tau}}$, we can derive
\begin{align} 
\bR_{Y_{\tau}^{s} Y_{\tau}^{s} } = \bH_{\tau}^{s} \bR_{X_{\tau} X_{\tau}} {\bH_{\tau}^{s}}^T +
\bR_{\zeta_{\tau}^{s}\zeta_{\tau}^{s}}.
\label{eq:meas-covariance}
\end{align}
Note that $\bR_{X_{\tau} X_{\tau}}$ is the covariance of the extended
Kalman smoother.

Next, we calculate the cross-covariance $\bR_{X_{k} Y_{\tau}^{s} }$.
Since $\bE\{ X_{\tau} \} = f_{\tau}(\mu_{X_{\tau-1}})$, we have for
any $\tau < k$:
\begin{align}
X_{k} - \bE\{ X_{k} \} &= \bF_{k}(X_{k-1} -
\mu_{X_{k-1}}) + \vartheta_{k} \\
&= \bF_{k}(\bF_{k-1}(X_{k-2} - \mu_{X_{k-2}}) + \vartheta_{k-1}) +
\vartheta_{k} \\
&= \bF_{k}\bF_{k-1}(\bF_{k-2}(X_{k-3} - \mu_{X_{k-3}}) +
\vartheta_{k-2}) +
\bF_{k}\vartheta_{k-1} + \vartheta_{k} \\
&= \bF_{k,\tau}(X_{\tau} - \mu_{X_{\tau}}) +
\sum_{j=\tau+1}^k \bF_{k,j} \vartheta_{j}
\end{align}
where we have introduced the notation $\bF_{k,\tau} =
\prod_{j=\tau+1}^k \bF_{j}$ and $\bF_{k,k} = \bI$. We can thus
evaluate the cross-covariance using the expression:
\begin{align}
\bR_{X_{k} Y_{\tau}^{s} } =\bF_{k,\tau} \bR_{X_{\tau} X_{\tau}} {\bH_{\tau}^{s}}^T.
\label{eq:crosscovariance}
\end{align}

\section{Asymptotic Optimality of the Proposed Algorithm} \label{Sec:5}

In this section we will consider the conditions under which the
unconditional measurement independence approximation made in the
previous section is expected to hold, assuming that the Gaussian
approximation is accurate. The assumption simplifies the algorithm
derivation and reduces its computational requirements, but it leads to
sub-optimality of the derived constrained MSE minimization
algorithm. The conditions established in this section help us
understand when the performance of the proposed sub-optimal algorithm
is expected to approach that of the optimal OOSM selection algorithm,
assuming that the Gaussian approximation and linearization are accurate.

The following theorem specifies that, under mild regularity
assumptions, if an asymptotic condition on the minimal eigenvalues of the noise matrices
holds, then the block-diagonal approximation employed to
derive the OOSM selection algorithm in the previous section holds
exactly. The proof is provided in Appendix~\ref{app:Proof_Theorem}.

\begin{theorem} \label{th:asymptotic_optimality} Let
  $\bR_{\mathcal{Z}_{k} \mathcal{Z}_{k}}$, $\bR_{X_{k}
    \mathcal{Z}_{k}}$ be defined as in \eqref{eq:covariances} and let
  $\bB_{\mathcal{Z}_{k} \mathcal{Z}_{k}}$ be the block-diagonal matrix
  whose diagonal blocks match those of $\bR_{\mathcal{Z}_{k}
    \mathcal{Z}_{k}}$ (the covariances of measurements from the same
  sensor at the same time).
Suppose that
  the following assumptions hold:
\begin{itemize}
\item[$\mathcal{A}_1$:] $\rho(\bR_{X_{n} X_{n}}) <
\infty$, $\forall k-\ell \leq n < k$
\item[$\mathcal{A}_2$:] $\rho(\bH_{m}^{s} {\bH_{m}^{s}}^T)^{1/2} < \infty$ and $\rho({\bH_{m}^{s}}^T \bH_{m}^{s})^{1/2} < \infty$, $\forall k-\ell \leq m
< k$ and $\forall s \in \mathcal{V}_m$
\item[$\mathcal{A}_3$:] $\rho(\bF_{m,n} \bF_{m,n}^T)^{1/2} < \infty$, $\forall k-\ell\leq n\leq m$ and $k-\ell \leq m \leq k$
\end{itemize}

Then we have for any $\ell,K < \infty$, $\mathcal{Z}_{k}$ and $k>1$:
\begin{align}
\min_{s,m}\lambda_{\min}(\bR_{\zeta_{m}^{s}\zeta_{m}^{s}})
\rightarrow \infty \Rightarrow | \tr \bR_{X_{k}
\mathcal{Z}_{k}}\bR_{\mathcal{Z}_{k}
\mathcal{Z}_{k}}^{-1}\bR_{\mathcal{Z}_{k} X_{k}} -  \tr \bR_{X_{k}
\mathcal{Z}_{k}}\bB_{\mathcal{Z}_{k}
\mathcal{Z}_{k}}^{-1}\bR_{\mathcal{Z}_{k} X_{k}} | \rightarrow 0,
\end{align}
where $\lambda_{\min}(\cdot) = \min_{i}\lambda_i(\cdot)$
\end{theorem}

The regularity conditions imposed in
Theorem~\ref{th:asymptotic_optimality} are mild and natural. Assumption
$\mathcal{A}_1$ requires the extended Kalman smoother covariance
$\bR_{X_{n} X_{n}}$ to have finite spectral radius. Thus assumption
$\mathcal{A}_1$ basically requests the stability (including the
numerical stability) of the EKS. Assumption $\mathcal{A}_2$ and
$\mathcal{A}_3$ require the spectral radia of matrices $\bH_{m}^{s}
{\bH_{m}^{s}}^T$, ${\bH_{m}^{s}}^T \bH_{m}^{s}$ and $\bF_{m,n}
\bF_{m,n}^T$ to be finite.  If the measurement
and transition functions, $h_{k}^{s}(\cdot)$ and $f_{k}(\cdot)$, are differentiable (sufficiently smooth),
leading to $\bF_{m}$ and $\bH_{m}^{s}$ with finite elements, then
assumptions $\mathcal{A}_2$ and $\mathcal{A}_3$ hold by the Gershgorin
disc theorem~\cite{Horn1990}. Any scenario when the EKS
functions normally and can be implemented leads to the assumptions
being satisfied.

The asymptotics in the theorem are with respect to
$\min_{s,m}\lambda_{\min}(\bR_{\zeta_{m}^{s}\zeta_{m}^{s}})
\rightarrow \infty$. The implications are best
illustrated by way of example. If
$\bR_{\zeta_{m}^{s}\zeta_{m}^{s}}$ is scalar for all sensors at all
times and $\bR_{\zeta_{m}^{s}\zeta_{m}^{s}} = {\sigma^2}_{m}^{s}$, then
$\lambda_{\min}(\bR_{\zeta_{m}^{s}\zeta_{m}^{s}}) =
{\sigma^2}_{m}^{s}$. The asymptotic condition thus implies that the
measurement noise variance approaches infinity for all sensors at all
times, or, equivalently, that all measurements become utterly
uninformative. If, on the other hand,
$\bR_{\zeta_{m}^{s}\zeta_{m}^{s}}$ is $2\times 2$ with equal
component variances, for all sensors at all times:
\begin{align}
\bR_{\zeta_{m}^{s}\zeta_{m}^{s}} = {\sigma^2}_{m}^{s}
\begin{pmatrix} 1 & r_{m}^{s} \\
r_{m}^{s} & 1
\end{pmatrix}
\end{align}
then $\lambda_{\min}(\bR_{\zeta_{m}^{s}\zeta_{m}^{s}}) =
{\sigma^2}_{m}^{s}(1-|r_{m}^{s}|)$. Thus the asymptotic
specifies that measurement components are not absolutely
(positively or negatively) correlated ($|r_{m}^{s}| \neq 1,
\forall m,s$), and that they have asymptotically large variance.

\section{Selective OOSM Particle Filter} \label{Sec:6}

In this section we specify an OOSM particle filter that employs the 
general OOSM selection framework presented in
Section~\ref{Sec:4}. For clarity, we describe the filter in the
context of a specific application scenario, but it can be easily
adapted to different delay models and OOSM processing costs. 

We consider a situation when there are several sensors sending
measurements (e.g. bearing or range) of the target to a common fusion
centre. All sensors are assumed to have communication issues leading
to OOSMs. An OOSM arrives at the fusion centre from a given sensor
with probability $p_{osm}$ and delay $d$. The delay $d$ is uniformly
distributed in the interval $[0,\ell]$. The probability $1-p_{osm}$
characterizes the events that (i) the OOSM is dropped in the network; reaches the fusion centre at
all. For example, this corresponds to the scenario when OOSMs delayed
by more than $\ell$ are automatically dropped by the network.  

We implement the proposed OOSM processing framework using the SEPF-EKS
algorithm of~\cite{Orguner2008}. In this case the OOSMs with the same
time stamp arriving from different sensors can be processed in one
sweep of SEPF-EKS algorithm (see Algorithm~\ref{alg:SEPF-EKS} and
associated discussion). Instead of a single OOSM $Z_{\tau,k}^{s}$ we
thus consider a set $\mathbf{Z}_{\tau,k}^{\{ \mathcal{I} \}}$
consisting of $\{Z_{\tau,k}^{s}\}$ and designated by the ordered index
set $\mathcal{I} = \{0,1\}^{|V_{\tau}|}$ such that $Z_{\tau,k}^{s} \in
\mathbf{Z}_{\tau,k}^{\{ \mathcal{I} \}}$ if and only if the element
corresponding to sensor s, 
$\mathcal{I}(s) = 1$.

We set the cost to process $\mathbf{Z}_{\tau,k}^{\{ \mathcal{I} \}}$
as $C_{\tau,k}^{\{ \mathcal{I} \}} = 1$, (the cost to run the
SEPF-EKS algorithm on a given hypothetical realization
$\mathbf{Z}_{\tau,k}^{\{ \mathcal{I} \}}$), irrespective of the
particular combination of $Z_{\tau,k}^{s}$. We make this choice
because the computational complexity of the SEPF-EKS algorithm is
approximately the same as one timestep of the particle filter. 
The average cost constraint analogous to~\eqref{eqn:cost_constraint} is then:
\begin{align}
\sum_{\mathcal{I} \in \mathfrak{I}} d_{\tau,k}^{\{ \mathcal{I} \}}
p_{\tau,k}^{\{\mathcal{I}\}} \leq C_{\ave},
\end{align}
where $d_{\tau,k}^{\{ \mathcal{I} \}}$ is the decision whether or not
to process a given realization $\mathbf{Z}_{\tau,k}^{\{ \mathcal{I}
  \}}$ and $\mathcal{J}$ is the set of all possible realizations of
$\mathcal{I}$. $C_{\ave}$ can be interpreted as the average number of
SEPF-EKS algorithm sweeps per filtering step or, in other words, the
average additional overhead caused by OOSM processing.

We have an expression for the MSE analogous to~\eqref{eqn:MSE_simplified}:
\begin{align}
\tr \var(X_{k}-\widehat X_{k}) = \tr\bR_{X_{k} X_{k} } -
\sum_{m=k-\ell}^{k-1} \sum_{\mathcal{I}\in \mathfrak{I}}
d_{\tau,k}^{\{\mathcal{I}\}} p_{\tau,k}^{\{\mathcal{I}\}}
R_{\tau,k}^{\{\mathcal{I}\}}.
\end{align}
Here $R_{\tau,k}^{\{\mathcal{I}\}}$ is calculated similarly
to~\eqref{eqn:measurement_utility} with $Y_{m}^{\{\mathcal{I}\}}$
being the vector constructed from those $Y_{m}^{s}$ for which
$\mathcal{I}(s)=1$:
\begin{align}
R_{m,k}^{\{\mathcal{I}\}} = \tr \bR_{X_{k} Y_{m}^{\{\mathcal{I}\}} }
\bR_{Y_{m}^{\{\mathcal{I}\}}Y_{m}^{\{\mathcal{I}\}} }^{-1}
\bR_{Y_{m}^{\{\mathcal{I}\}} X_{k} }.
\end{align}
The probability $p_{\tau,k}^{\{\mathcal{I}\}}$ that an OOSM with a given
sensor combination $\mathcal{I}$ active at time $\tau$ arrives at
time $k$ can be calculated as:
\begin{align}
p_{\tau,k}^{\{\mathcal{I}\}} = \prod_{s \in \mathcal{V}_{\tau} :
\mathcal{I}(s)=1} p_{\tau,k}^{s} \prod_{j \in \mathcal{V}_{\tau} :
\mathcal{I}(j)=0} (1-p_{\tau,k}^{j}).
\end{align}
Here $p_{\tau,k}^{s} = 0$ if the measurements from sensor $s$ at time
$\tau$ have already arrived. If not, then:  
\begin{align}
p_{\tau,k}^{s} = \Pr\{ \Delta_{\tau}^{s} = k-\tau | b_{\tau,k-1}^{s} = 0,\ldots, b_{\tau,\tau}^{s}=0
\}
\end{align}
where $\Delta_{\tau}^{s}$ is the delay that the OOSM from sensor $s$
experiences at time $\tau$.  For the case of the uniform delay
distribution and probability of successful transmission $p_{osm}$, we have:
\begin{align}
p_{\tau,k}^{s} = \frac{p_{osm}}{\ell+1-(k-\tau)}
\end{align}

Equipped with the expressions above we can calculate $\widetilde
R_{\tau,k}^{\{ \mathcal{I} \}}$, the analog of $\widetilde
R_{\tau,k}^{s}$, and apply a slightly modified version of
Algorithm~\ref{alg:th_select} to set the threshold $\gamma_k$.  This
algorithm employs a similar measurement covariance matrix
block-diagonality approximation as in the general framework described
in Section~\ref{Sec:4}. In this case, however, the blocks are larger
and consist of matrices
$\bR_{Y_{m}^{\{\mathcal{I}\}}Y_{m}^{\{\mathcal{I}\}}}$, rather than
$\bR_{Y_{m}^{s}Y_{m}^{s}}$. The modified approximation is thus
that blocks $\bR_{Y_{m}^{\{\mathcal{I}\}}Y_{n}^{\{\mathcal{J}\}}}$ are
close to zero for all combinations of sensors $\mathcal{I}$ and
$\mathcal{J}$ and any $m \neq n$.

As a final heuristic refinement of the algorithm we use the
OOSM-GARP algorithm to process those OOSMs for which the SEPF-EKS
algorithm performs poorly, namely the highly informative measurements
that should induce significant shifts in the current filtering
distribution. We add a test to check whether the effective number of
samples in the particle filter drops significantly after the
application of the SEPF-EKS processing; if this occurs, we reprocess
the OOSM using the OOSM-GARP filter. This allows the algorithm to
adjust both weights and \emph{locations} of particles to
account for the new information embedded in the OOSMs. We have
observed that this step greatly improves the performance of the
filter in difficult situations at a minimal cost.

The OOSM particle filtering algorithm based on the above discussion is
presented in Algorithm~\ref{alg:PROPOSED}. This algorithm describes
only the OOSM processing procedure corresponding to \verb,ProcessOOSM,
in Algorithm~\ref{alg:PFOOSM} (Algorithm~\ref{alg:PFOOSM} presents the
complete high level OOSM particle filter pseudocode). In
Algorithm~\ref{alg:PROPOSED}, as we discussed in Section~\ref{Sec:4},
we first calculate the sequence of EKS means and covariance matrices,
which are further used to compute the Jacobians and the utilities
$\{\widetilde{R}_{\tau, k}\}$. These are used in \verb,CalcGamma, (a
minor modification of Algorithm~\ref{alg:th_select}), which has the
task of setting the current value of threshold $\gamma_k$. This
threshold is used to determine which OOSMs should be processed with
the function \verb,ProcessSEPF-EKS, summarized in
Algorithm~\ref{alg:SEPF-EKS}. The failure of this algorithm, which is
expressed through particle degeneracy, is detected via the second
threshold test (where the value of $\nu$ should be small,
e.g. $1/40$). If a failure is detected, the algorithm switches to
recalculate the current particle set via function
\verb,ProcessOOSM-GARP, summarized in
Algorithm~\ref{alg:ProcessOOSM-GARP}.

\begin{algorithm}[H] \label{alg:PROPOSED}

\SetKwFunction{ProcessOOSM}{ProcessOOSM}
\SetKwFunction{ParticleFilter}{ParticleFilter}
\SetKwFunction{SaveGauss}{SaveGauss}
\SetKwFunction{CalcGamma}{CalcGamma} \SetKwFunction{EKS}{EKS}
\SetKwFunction{ProcessSEPF}{ProcessSEPF-EKS}
\SetKwFunction{ProcessGARP}{ProcessOOSM-GARP}

At time $k$

\KwIn{ $\mathcal{Z}_k$, $\bxi_{k}$, $\bomega_{k}$, $\bOmega_k$, $C_{\ave}$} 

($\bmu_{X_{k-\ell:k}}$, $\bR_{X_{k-\ell:k} X_{k-\ell:k}}$)
$\leftarrow$ \EKS{$\bOmega_k$} \;

($\gamma_{k}$, $\{\widetilde{R}_{\tau, k}\}$) $\leftarrow$
        \CalcGamma{$\bmu_{X_{k-\ell:k}}$, $\bR_{X_{k-\ell:k} X_{k-\ell:k}}$, $C_{\ave}$} \;

EKSfailed = 0 \;

\For{$\tau : \mathbf{Z}_{\tau, k} \in \mathcal{Z}_k$}{

    \If{$\widetilde{R}_{\tau, k} \geq \gamma_{k}$}{

        ($N_{\prior}$) $\leftarrow$ $1 / \| \bomega_{k} \|_2^2$ \;

        ($\bomega_{k}$) $\leftarrow$
        \ProcessSEPF{$\mathbf{Z}_{\tau, k}$, $\bxi_{k}$, $\bomega_{k}$, $\bOmega_k$} \;

        ($N_{\post}$) $\leftarrow$ $1 / \| \bomega_{k} \|_2^2$ \;

        \If{$N_{\post} < \nu N_{\prior}$}{

            EKSfailed = 1 \;

            break \;

        }

    }

}

\uIf{EKSfailed}{

($\bomega_{k}$, $\bxi_{k}$, $\bOmega_k$) $\leftarrow$
        \ProcessGARP{$\mathcal{Z}_k$, $\bOmega_k$} \;

}

\caption{Particle Filter with selective OOSM processing
(ProcessOOSM)}
\end{algorithm}

\section{Numerical Experiments} \label{Sec:7}

\noindent In our simulations we consider a two-dimensional scenario
with a single target that makes a clockwise coordinated turn of
radius $500m$ with a constant speed $200km/h$. It starts in
the y-direction with initial position $[-500m,500m]$ and is tracked for
$40$ seconds.

The target motion is  modeled in the filters by the nearly
coordinated turn model~\cite{Bar-Shalom2001} with unknown constant
turn rate and cartesian velocity. The state of the target is given
as $x_k=[p_k^x,p_k^y, v_k^x,v_k^y, \omega_k]^T $, where $p,v$ and
$\omega$ denote the position, velocity and turn rate respectively.
The dynamic model for the coordinated turn model is
\begin{equation*}
X_{k+1}=
\begin{pmatrix}
1 & 0 & \frac{\sin(\omega_k)}{\omega_k}    & \frac{\cos(\omega_k)-1}{\omega_k}    & 0 \\
0 & 1 & \frac{1- \cos(\omega_k)}{\omega_k} & \frac{\sin(\omega_k)}{\omega_k}      & 0 \\
0 & 0 & \cos(\omega_k)                     & -\sin(\omega_k)                      & 0 \\
0 & 0 & \sin(\omega_k)                     & \cos(\omega_k)                       & 0 \\
0 & 0 & 0                                           & 0 & 1
\end{pmatrix}
X_k + \vartheta_{k+1}
\end{equation*}
where $\vartheta_{k+1}$ is Gaussian process noise,
$\vartheta_{k+1}\sim \mathcal{N} (0,\bV_{k+1})$,
$\bV_{k+1}=\diag([30^2, 30^2, 10^2,10^2,0.1^2 ])$, and the sampling
period is $1$ second. We assume that the filter initially knows
little about the state of the target and therefore it is initialized
with the state value $\mu_{X_0} = [0,0,0,0,0]^T$ and a large
covariance $\bR_{X_0 X_0}= \diag([1000^2,1000^2,30^2,30^2,0.1^2])$.

There are three sensors S1, S2 and S3 sending bearing-only
measurements of the target to a common fusion centre. The sensor
locations are $[S_1^x,S_1^y]=[-200,0]$, $[S_2^x,S_2^y]=[200,0]$,
$[S_3^x,S_3^y]=[-750,750]$ and the bearings-only measurement
function is:
\begin{equation}
\mathit{h}_k^j(x_k)=\arctan(\frac{p_k^y-S_j^y}{p_k^x-S_j^x}) \quad\quad
j=1,2,3.
\end{equation}
The measurements from the sensors are corrupted with additive
independent Gaussian noises with zero mean and standard deviation
$\sigma_s=0.05$. An OOSM arrives at the fusion centre from a given
sensor with probability $p_{osm}$ and delay $d$. The delay $d$ is
uniformly distributed in the interval $[0,5]$. The probability
$p_{osm}$ that an OOSM reaches the fusion centre at all is set to
$0.7$.

\begin{figure}[htp]
\centering \subfigure[Position RMSE versus time]{ \label{fig:RMS:2000}
\includegraphics[width = 11cm]{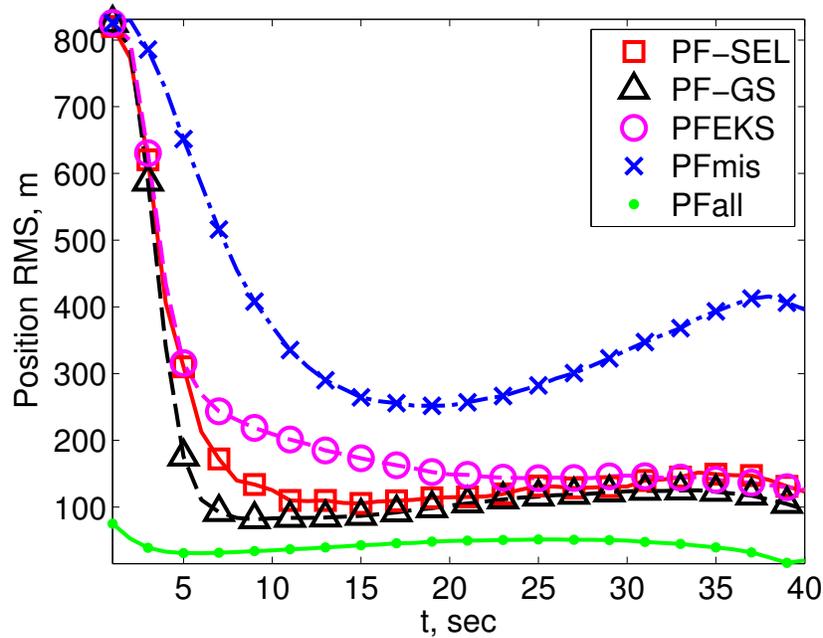}}
 \subfigure[Boxplots of position RMSEs versus time (seconds)]{\label{fig:Errorbar} 
\includegraphics[width = 15cm]{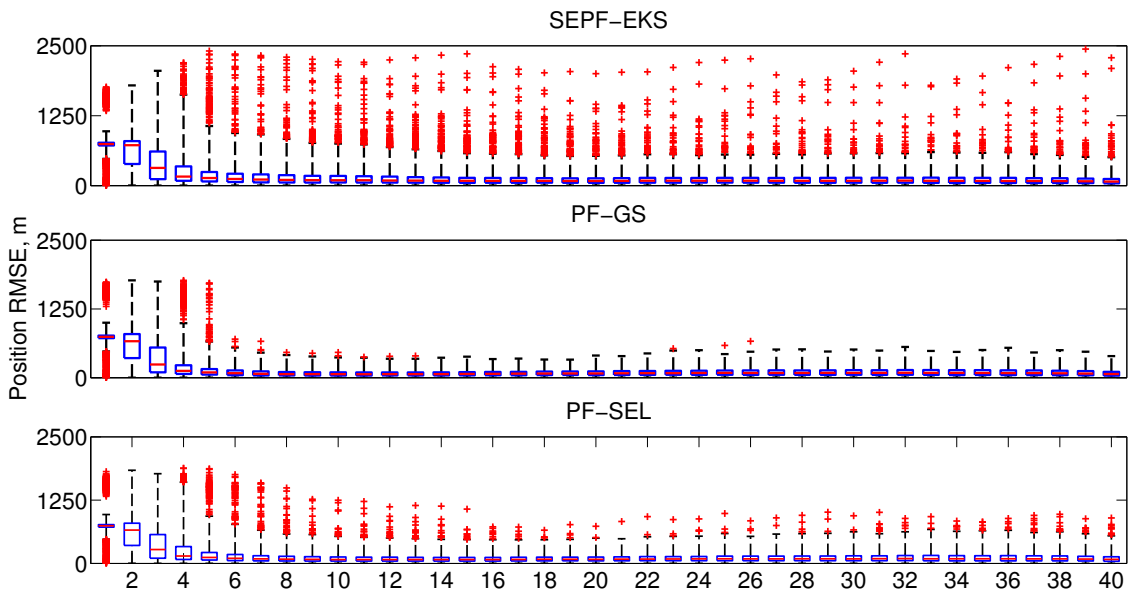}}
\caption{Tracking performance of the particle filters as a function
of time using RMS error as a performance metric. (a) The curves show the
  means of 5000 Monte-Carlo trials. (b) Errorbars showing the variation of position RMS for
  \textit{SEPF-EKS}, \textit{PF-GS} and \textit{PF-SEL}, when they use 2000
  particles. The box has lines at the lower quartile, median(red line), and upper quartile values. Outliers (red '+') are values beyond the range of 5 times the interquartile range from the ends of the box. }
\label{fig:RMS} 
\end{figure}

\subsection {Benchmarked Filters}
\noindent We have implemented five different particle filters, all
based on the Sampling Importance Resampling (SIR) filtering
paradigm~\cite{Gordon1993}. The prior distribution is used as the
importance function\footnote{Although better performance could be
  achieved by using a more carefully-chosen importance function, this
  generally comes at the cost of some computational expense. By using
  the same, simple importance function for all particle filters we
  achieve a fair performance comparison.}. The filters were implemented in Matlab and the
code was highly optimized.

\textit{PFall}: collects all measurements from all active sensors
(no OOSMs). This is an idealized filter that provides a performance
benchmark; a real-time implementation is impossible.

\textit{PFmis}: discards all OOSMs and therefore only processes the
measurements with zero delay.

\textit{SEPF-EKS}: Storage efficient particle filter using EKS
smoothing as described in~\cite{Orguner2008} (Algorithm~\ref{alg:SEPF-EKS}).

\textit{PF-GS}: The OOSM-GARP algorithm described in
Algorithm~\ref{alg:ProcessOOSM-GARP}.

\textit{PF-SEL}: Selective OOSM processing based on the proposed
framework and described in Algorithm~\ref{alg:PROPOSED}.

We use the root mean-squared (RMS) position error to compare the
performances of the particle filters. Let $({p_k^x},{p_k^y})$ and
$(\hat{p}_{k,i}^{x},\hat{p}_{k,i}^{y})$ denote the true and
estimated target positions at time step $k$ for the $i$-th of $M$
Monte-Carlo runs. The RMS position error at $k$ is calculated as
\begin{align}
\RMS_k = \sqrt{\frac{1}{M} \sum_{i=1}^{M}
(\hat{p}_{k,i}^{x}-{p_k^x})^2 + (\hat{p}_{k,i}^{y}-{p_k^y})^2 }
\end{align}

\subsection{Results and Discussion}

\noindent In our first experiment we fix the computational cost
$C_{\ave} = 0.6$. We thus allow $0.6$ sweeps of the SEPF-EKS
algorithm to be performed on average per filtering step. Over an
extended period of time of sufficiently large length $L$ this leads
to an additional OOSM processing overhead of $\sim 0.6 L C_{EKS}$
where $C_{EKS}$ is the cost of one sweep of the SEPF-EKS algorithm.
If we do not apply the proposed procedure and process all the
available OOSMs this cost in our application scenario is approximately
$1.5 L C_{EKS}$. This implies that we process only
approximately $40\%$ of all measurements.

In Fig.~\ref{fig:RMS}, we plot the respective RMS position performance
for the tracking period of $40s$ for the algorithms with these
settings. Corresponding error-bar plots of the RMS performance are
shown in Fig.~\ref{fig:Errorbar}. The actual number of individual
OOSMs processed by the SEPF-EKS after application of the first
threshold $\gamma_k$ measured in our experiment is $40.04\%$. After
the second threshold $\nu = 1/40$ the percentage of most informative OOSMs
processed by rerunning the particle filter using OOSM-GARP is $1.57\%$.

Fig.~\ref{fig:RMS} indicates that despite processing only a
relatively small fraction of the OOSMs, the proposed algorithm
performs almost as well as the much more complex OOSM-GARP algorithm
(\textit{PF-GS}). The calculation of the selection criterion has
minimal overhead, so discarding the uninformative measurements
results in significant computational savings. Thus the proposed
filter is more computationally efficient than the SEPF-EKS filter
and yet, as can be seen from Fig.~\ref{fig:RMS}, it has better RMS
performance. Fig.~\ref{fig:Errorbar} indicates that the performance
of SEPF-EKS is not as stable as that of \textit{PF-GS} and
\textit{PF-SEL}. In the proposed algorithm the increased robustness
and performance stability is achieved by using the second threshold
to detect situations when reweighting particles induces sample degeneracy
problems.

\begin{figure*}[t]
\centering \subfigure[$N=2000$, $t=10$]{
\label{fig:RMS_Complexity:N2000_t10}
\includegraphics[width = 8cm]{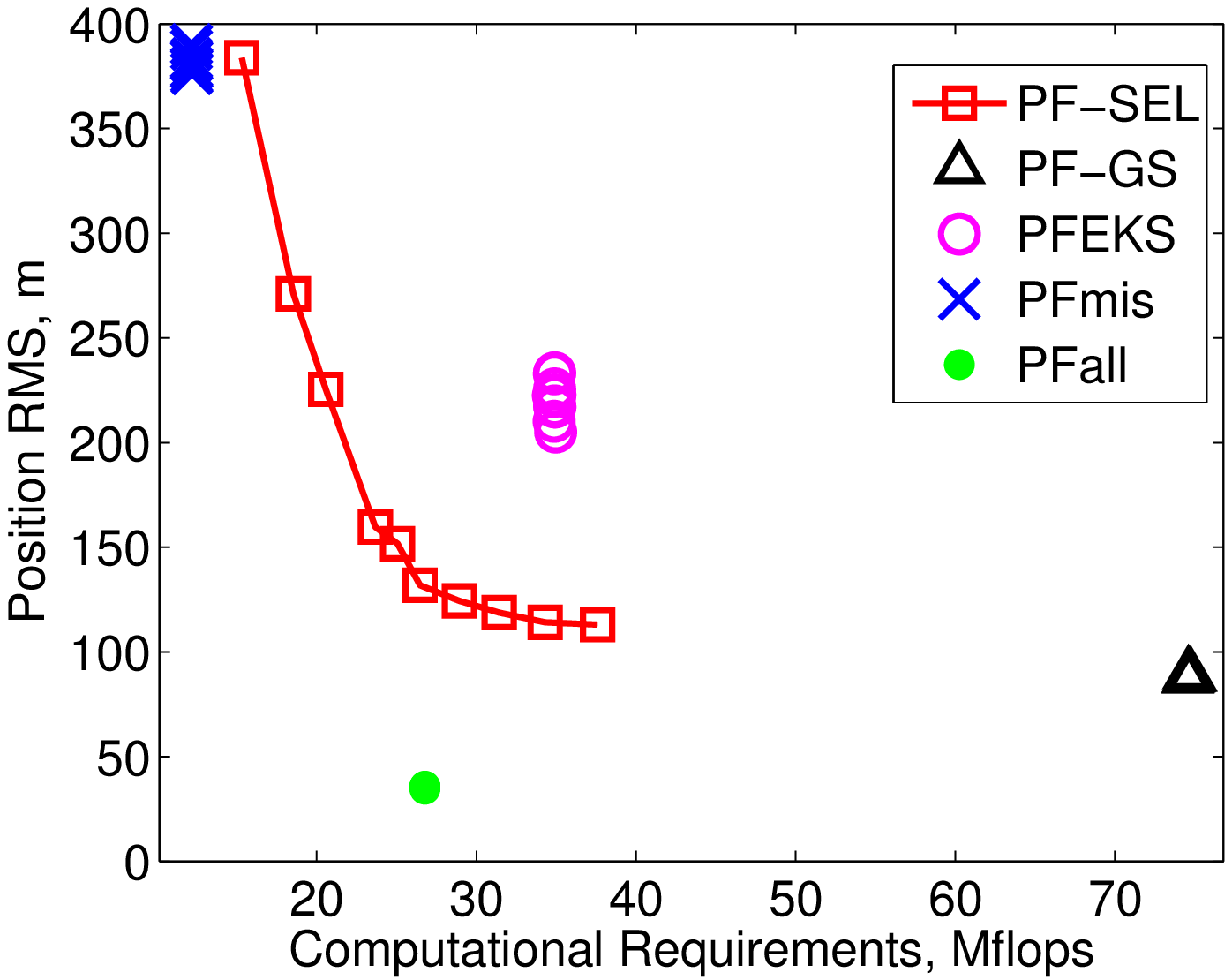}}
\subfigure[$N=2000$, $t=20$]{\label{fig:RMS_Complexity:N2000_t20} 
\includegraphics[width = 8cm]{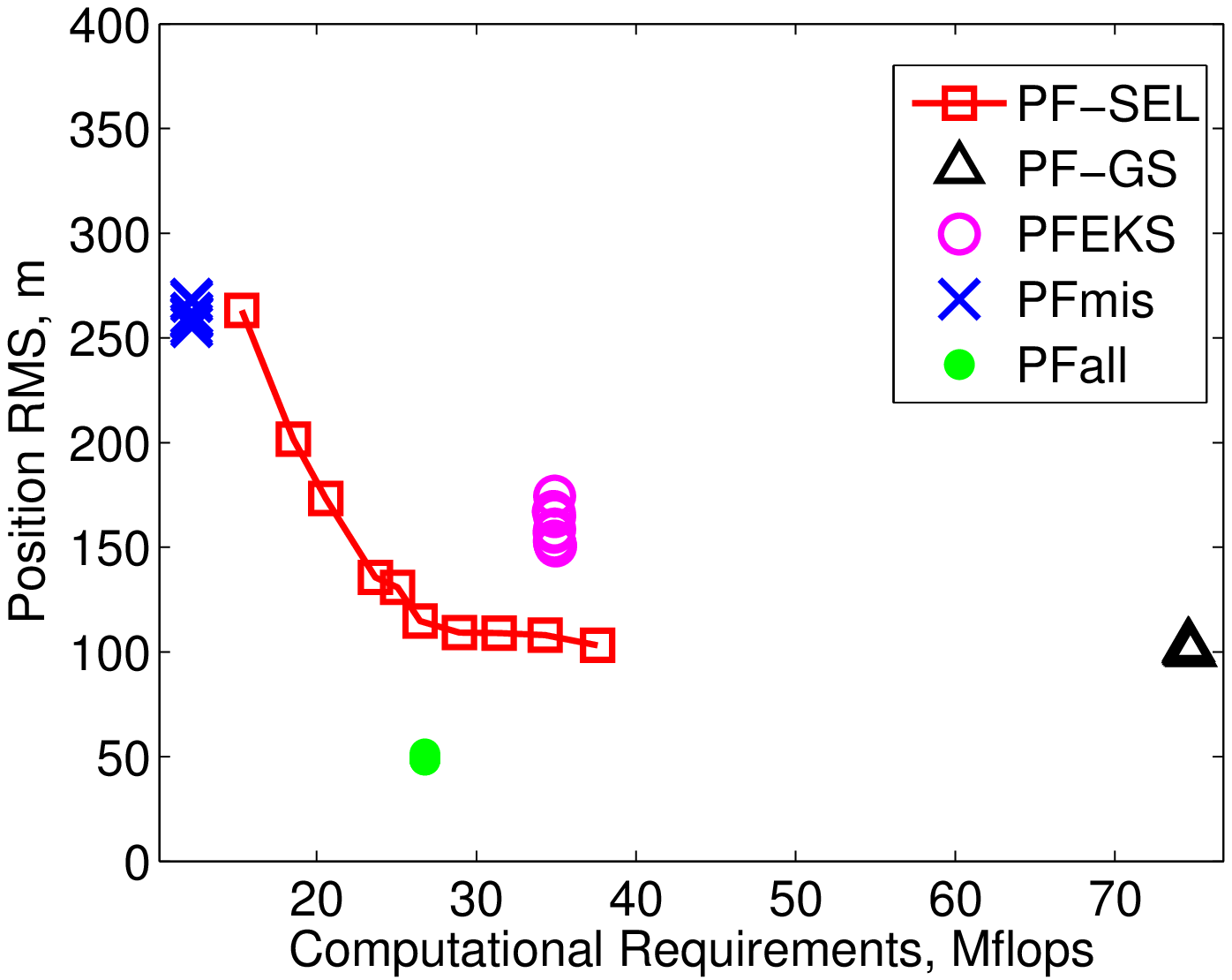}}
\subfigure[$N=2000$, $t=30$]{\label{fig:RMS_Complexity:N2000_t30} 
\includegraphics[width = 8cm]{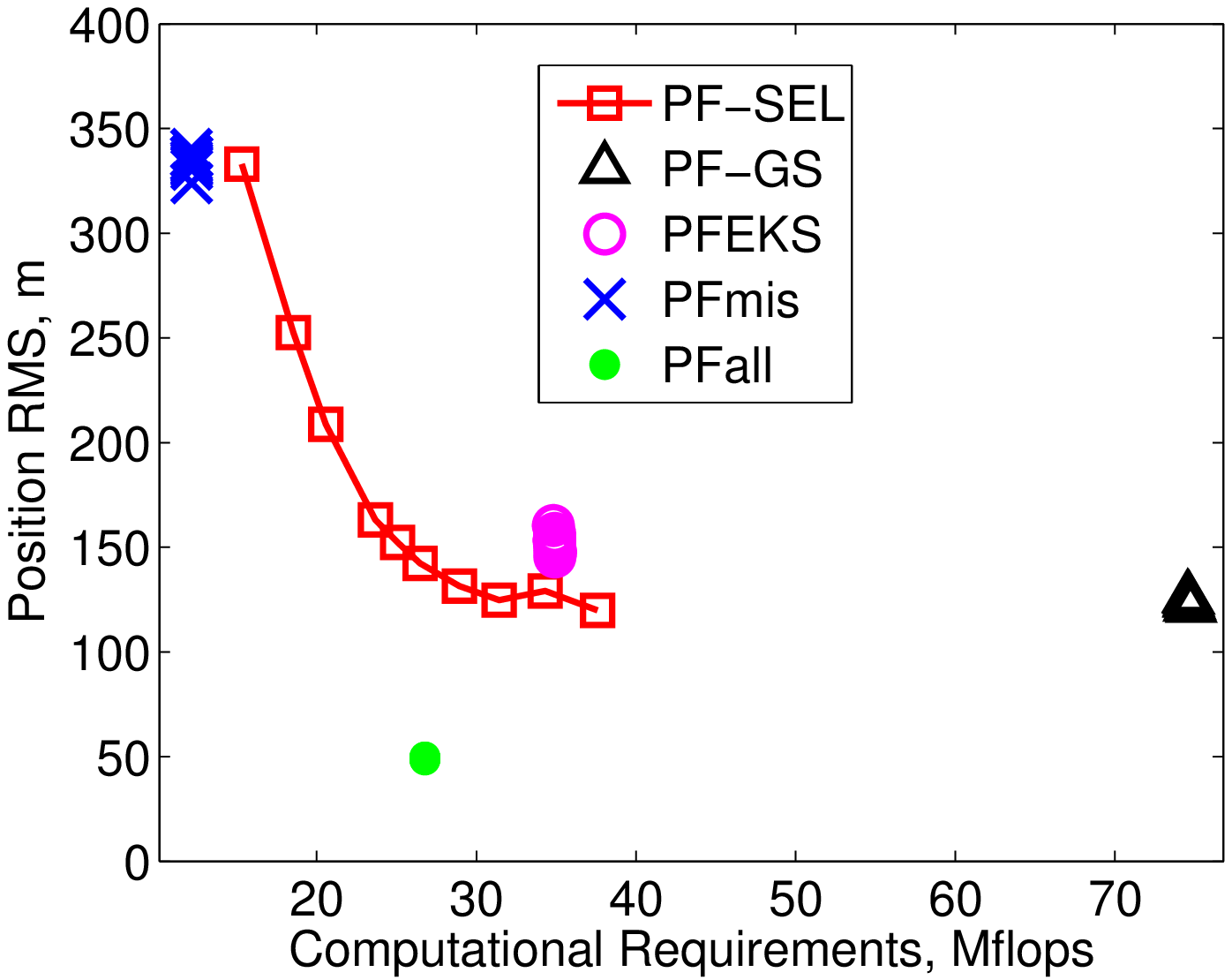}}
\caption{RMS vs Complexity from $10$ simulations with different
$C_{\ave}$. Each simulation shows the average of $1000$ MC runs.We
select three timesteps, $t=10,20,30$ for filters with $2000$ and
$5000$ particles. The complexity is measured by running time for
tracking $40$s of each filter. The results are run on a Dell laptop
with Genuine Intel(R) CPU T2400 1.83GHz, 0.99GB RAM and Win-XP OS.}
\label{fig:RMS_Complexity} 
\end{figure*}

In our next experiment we study the computational complexity versus
accuracy trade-off for the proposed algorithm. We illustrate this by
varying the computational complexity of the proposed algorithm by
adjusting $C_{\ave}$ and plotting the RMS error vs. computational load
measured in MATLAB. We use the following values to control the OOSM
processing overhead: $C_{\ave} = \{0, 0.1, 0.2, 0.4, 0.5, 0.6, 0.8, 1,
1.3, 2\}$, $\gamma_2= 1/40$. These results are reported in
Fig.~\ref{fig:RMS_Complexity}. In this figure we show the relationship
between complexity and performance for the proposed algorithm with ten
values of $C_{\ave}$ and results of 10 simulations for other
algorithms. Each simulation involves 1000 Monte Carlo runs. We compare
the performance of all particle filters when they use $2000$
particles; qualitatively similar results were observed for $1000$ and
$5000$ particles. When the thresholds are chosen so that the proposed
filter has the same computational complexity as SEPF-EKS, it achieves
significantly better tracking performance. Alternatively, for the
same fixed RMS error performance, the selective processing algorithm
reduces the computational load by $30-40\%$. Compared to the OOSM-GARP
algorithm, a $50\%$ reduction in computational requirements leads to
only a small increase in estimation error. The results illustrate that
we can adjust $C_{\ave}$ to control the trade-off between the average
computational load or power supply consumption and the tracking
performance.

\section{Conclusions}\label{Sec:8}

\noindent This paper presents a framework for selective processing
of the out-of-sequence measurements. Based on this framework we
develop a computationally efficient algorithm for delay-tolerant
particle filtering that has limited memory requirements. By
identifying and discarding the uninformative delayed measurements,
the algorithm reduces the computational requirements. By processing
the most informative measurements with a re-run particle filter, the
algorithm achieves better tracking performance than the storage
efficient particle filter of~\cite{Orguner2008}.

In our framework, the threshold to discard uninformative measurements
is set by minimizing the one-step MSE calculated from the Gaussian
approximation of posterior at every filtering time instant. The
threshold setting could be improved by employing a finite
horizon dynamic programming technique to take into account the MSE
reduction over several forthcoming steps. It is also interesting to
explore whether the fusion centre can provide feedback to the sensor
nodes so that they can locally assess measurement informativeness.
This would allow sensor nodes to avoid unnecessary energy
expenditure by discarding uninformative measurements prior to
transmission.

\appendices
\section{Proof of Theorem 1} \label{app:Proof_Theorem}

We here provide a proof of Theorem 1. 

We first state a lemma that is employed within the main proof. Denote
the spectral radius of a matrix by $\rho(\cdot) = \max_i
|\lambda_{i}(\cdot)|$. The proof of the lemma involves expanding the
variational characterization of the spectral radius in terms of the
blocks of $\bA$ and applying the Cauchy-Schwarz inequality to each
term in the expansion.
\begin{lemma} \label{Lemma_A} Let $\bA \in \mathbb{R}^{K\times K}$ be a block matrix
  consisting of blocks $\bA_{i,j}\in \mathbb{R}^{K_i\times M_j}$. Then
  $\rho(\bA) \leq \sum_{i,j} \rho(\bA_{i,j}^T\bA_{i,j})^{1/2}$.
\end{lemma}
\begin{IEEEproof}
Let $\{\bx_{i}\}$ and $\{ \by_{j} \}$ be two partitionings of vector
$\bX \in \mathbb{R}^{K}$ such that $\bX^T = [\bx_{1}^T, \bx_{2}^T,
\ldots]$, $\bx_{i} \in \mathbb{R}^{K_i}, \forall i$ and $\bX^T =
[\by_{1}^T, \by_{2}^T, \ldots]$, $\by_{j} \in \mathbb{R}^{M_j},
\forall j$. Write the variational characterization of the spectral
radius and expand it in terms of blocks:
\begin{align} \label{eqn:rho_bA_decomp}
\rho(\bA) &= \max_{\bX \neq 0} \frac{|\bX^T \bA \bX|}{\bX^T \bX} \\
&= \max_{\bX \neq 0} \frac{|\sum_{i,j} \bx_i^T \bA_{i,j}
\by_{j}|}{\bX^T \bX} \\
&\leq \max_{\bX \neq 0} \frac{\sum_{i,j} |\bx_i^T \bA_{i,j}
\by_{j}|}{\bX^T \bX} \\
&\leq \sum_{i,j} \max_{\bX \neq 0} \frac{ |\bx_i^T \bA_{i,j}
\by_{j}|}{\bX^T \bX}.
\end{align}
Now, for every summand use the Cauchy-Schwarz inequality and take
into account the fact that $\bx_i^T \bx_i /\bX^T \bX \leq 1, \forall
i$:
\begin{align}
\max_{\bX \neq 0} \frac{ |\bx_i^T \bA_{i,j} \by_{j}|}{\bX^T \bX}
&\leq \max_{\bX \neq 0} \frac{ |\bx_i^T \bx_i|^{1/2}}{|\bX^T
\bX|^{1/2}} \frac{ |\by_{j}^T \bA_{i,j}^T \bA_{i,j}
\by_{j}|^{1/2}}{|\bX^T \bX|^{1/2}} \\
&\leq \max_{\bX \neq 0} \frac{ |\by_{j}^T \bA_{i,j}^T \bA_{i,j}
\by_{j}|^{1/2}}{|\bX^T \bX|^{1/2}}.
\end{align}
Similarly, note the fact that $\by_j^T \by_j \leq \bX^T \bX, \forall
j$:
\begin{align}
\max_{\bX \neq 0} \frac{ |\by_{j}^T \bA_{i,j}^T \bA_{i,j}
\by_{j}|^{1/2}}{|\bX^T \bX|^{1/2}} &\leq \max_{\bX \neq 0} \frac{
|\by_{j}^T \bA_{i,j}^T \bA_{i,j} \by_{j}|^{1/2}}{|\by_{j}^T
\by_{j}|^{1/2}} \\
&= \left( \max_{\by_{j} \neq 0} \frac{ |\by_{j}^T \bA_{i,j}^T
\bA_{i,j} \by_{j}|}{|\by_{j}^T \by_{j}|} \right)^{1/2}.
\end{align}
This is the variational characterization of $\rho(\bA_{i,j}^T
\bA_{i,j})^{1/2}$. Substituting it into~\eqref{eqn:rho_bA_decomp}
completes the proof.
\end{IEEEproof}

\begin{IEEEproof}[Proof of Theorem 1]

Employing the EKF linear approximation in \eqref{eqn:state_space_lin}
and using the independence of measurement and diffusion noises from each other
and from the state and independence of $\zeta_{m}^{s}$ and
$\zeta_{n}^{j}$ for any $m\neq n$ or $s\neq j$, we have for $m>n$
and any $s,j$:
\begin{align} \label{eqn:meas_cross_cov}
\bR_{Y_{m}^{s} Y_{n}^{j} } &= \bH_{m}^{s} \bF_{m,n} \bR_{X_{n} X_{n}} {\bH_{n}^{j}}^T
\end{align}
Note that for $m<n$ $\bR_{Y_{m}^{s} Y_{n}^{j} } = \bR_{Y_{n}^{s}
  Y_{m}^{j} }^T = \bH_{m}^{s} \bR_{X_{m} X_{m}} \bF_{n,m}^T
{\bH_{n}^{j}}^T$.

Recall that $\bB_{\mathcal{Z}_{k}\mathcal{Z}_{k}}$ is the
block-diagonal matrix whose blocks match the diagonal blocks of
$\bR_{\mathcal{Z}_{k} \mathcal{Z}_{k}}$. We now establish finite upper
bounds on two spectral radii, $\rho( \bB_{\mathcal{Z}_{k}
  \mathcal{Z}_{k}} - \bR_{\mathcal{Z}_{k}\mathcal{Z}_{k}})$ and
$\rho(\bR_{X_{k}\mathcal{Z}_{k}}\bR_{\mathcal{Z}_{k}X_{k}})$.
Throughout the proof we employ the fact that
$\rho(\bC^T\bA\bC)\leq\rho(\bA)\rho(\bC^T\bC)$ for a square matrix
$\bA$ and an arbitrary real matrix $\bC$. This follows from the
variational characterization of spectral radius:
\begin{align}
\rho(\bC^T \bA \bC) &= \max_{\bx\neq 0}\frac{|\bx^T\bC^T \bA \bC
\bx|}{\bx^T\bx} = \max_{\bx\neq 0}\left(\frac{|\bx^T\bC^T \bA \bC
\bx|}{\bx^T\bC^T\bC\bx} \frac{\bx^T\bC^T\bC\bx}{\bx^T\bx}\right) \\
&\leq \max_{\bx\neq 0}\left( \max_{\by\neq 0}
\frac{|\by^T\bA\by|}{\by^T\by}
\frac{\bx^T\bC^T\bC\bx}{\bx^T\bx}\right) \\
&\leq \rho(\bA) \rho(\bC^T \bC),
\end{align}

Since all the diagonal blocks of $\bB_{\mathcal{Z}_{k}
  \mathcal{Z}_{k}} - \bR_{\mathcal{Z}_{k}\mathcal{Z}_{k}}$ are zero,
we have from Lemma~\ref{Lemma_A} and \eqref{eqn:meas_cross_cov}:
\begin{align} \label{eqn:technical_2}
\rho( &\bB_{\mathcal{Z}_{k} \mathcal{Z}_{k}} - \bR_{\mathcal{Z}_{k}
\mathcal{Z}_{k}}) \nonumber \\
&\leq
\sum_{m=k-\ell}^{k-1}\sum_{s\in\mathcal{S}_{m,k}} \left(
\sum_{n=k-\ell, n\neq m}^{k-1} \sum_{j\in\mathcal{S}_{n,k}}
\rho(\bR_{Y_{m}^{s}Y_{n}^{j}} \bR_{Y_{n}^{j}Y_{m}^{s}})^{1/2} +
\sum_{j\in\mathcal{S}_{m,k},j\neq
s} \rho(\bR_{Y_{m}^{s}Y_{m}^{j}} \bR_{Y_{m}^{j}Y_{m}^{s}})^{1/2}\right) \nonumber\\
&\leq K\ell(K\ell-1) \max_{s,m} \max_{j\neq s \vee n\neq m}
\rho(\bR_{Y_{m}^{s}Y_{n}^{j}}
\bR_{Y_{n}^{j}Y_{m}^{s}})^{1/2} \nonumber\\
&\leq K\ell(K\ell-1) \max_{s,m} \max_{j\neq s \vee n\neq m}
\max[\rho(\bH_{m}^{s} \bF_{m,n} \bR_{X_{n} X_{n}} {\bH_{n}^{j}}^T
{\bH_{n}^{j}} \bR_{X_{n}
X_{n}} \bF_{m,n}^T {\bH_{m}^{s}}^T)^{1/2} , \nonumber\\
& \quad\quad\quad\rho(\bH_{n}^{j} \bF_{n,m} \bR_{X_{m} X_{m}} {\bH_{m}^{s}}^T
{\bH_{m}^{j}} \bR_{X_{m}
X_{m}} \bF_{n,m}^T {\bH_{n}^{j}}^T)^{1/2}] \nonumber\\
&\leq K\ell(K\ell-1) \max_n \rho(\bR_{X_{n} X_{n}}) \max_{s,m}
\rho(\bH_{m}^{s} {\bH_{m}^{s}}^T)^{1/2} \max_{s,m}
\rho({\bH_{m}^{s}}^T \bH_{m}^{s})^{1/2}  \max_{n \leq m}
\rho(\bF_{m,n}\bF_{m,n}^T)^{1/2}
\end{align}
Observing that $\bR_{X_{k}
\mathcal{Z}_{k}}\bR_{\mathcal{Z}_{k}X_{k}} = \sum_{s,m} \bR_{X_{k}
Y_{m}^{s}} \bR_{X_{k} Y_{m}^{s}}^T$ and
recalling~\eqref{eq:crosscovariance} we can write:
\begin{align} \label{eqn:technical_1}
\rho(\bR_{X_{k} \mathcal{Z}_{k}}\bR_{\mathcal{Z}_{k}X_{k}}) &\leq
\sum_{s,m} \rho(\bR_{X_{k} Y_{m}^{s}} \bR_{X_{k}
Y_{m}^{s}}^T) \\
&\leq K\ell \max_{s,m} \rho(\bF_{k,m} \bF_{k,m}^T)
\rho({\bH_{m}^{s}}^T {\bH_{m}^{s}}) \rho(\bR_{X_{m} X_{m}})^2.
\end{align}
Assumptions $\mathcal{A}_1$-$\mathcal{A}_3$ ensure that the bounds
in~\eqref{eqn:technical_2} and \eqref{eqn:technical_1} are finite.

We now develop an upper bound for $\rho(\bR_{\mathcal{Z}_{k} \mathcal{Z}_{k}}^{-1}-\bB_{\mathcal{Z}_{k}
\mathcal{Z}_{k}}^{-1})$
\begin{align} \label{eqn:technical_4}
\rho(\bR_{\mathcal{Z}_{k} \mathcal{Z}_{k}}^{-1}-\bB_{\mathcal{Z}_{k}
\mathcal{Z}_{k}}^{-1}) &=\rho(\bR_{\mathcal{Z}_{k} \mathcal{Z}_{k}}^{-1}(
\bB_{\mathcal{Z}_{k} \mathcal{Z}_{k}} - \bR_{\mathcal{Z}_{k}
\mathcal{Z}_{k}}) \bB_{\mathcal{Z}_{k} \mathcal{Z}_{k}}^{-1})\nonumber
\\&\leq \rho(\bR_{\mathcal{Z}_{k}
\mathcal{Z}_{k}}^{-1})\rho( \bB_{\mathcal{Z}_{k} \mathcal{Z}_{k}} -
\bR_{\mathcal{Z}_{k} \mathcal{Z}_{k}}) \rho(\bB_{\mathcal{Z}_{k}
\mathcal{Z}_{k}}^{-1}) \nonumber\\
&= \lambda_{\min}^{-1}(\bR_{\mathcal{Z}_{k} \mathcal{Z}_{k}})
\lambda_{\min}^{-1}(\bB_{\mathcal{Z}_{k} \mathcal{Z}_{k}}) \rho(
\bB_{\mathcal{Z}_{k} \mathcal{Z}_{k}} - \bR_{\mathcal{Z}_{k}
\mathcal{Z}_{k}}). 
\end{align}

Since the eigenvalues of the block-diagonal matrix are the
eigenvalues of its blocks we have:
$\lambda_{\min}(\bB_{\mathcal{Z}_{k} \mathcal{Z}_{k}}) =
\min_{s,m}\lambda_{\min}(\bR_{Y_{m}^{s}Y_{m}^{s}})$. This implies:
\begin{align}
\lambda_{\min}(\bB_{\mathcal{Z}_{k} \mathcal{Z}_{k}}) 
&=\min_{s,m}\lambda_{\min}(\bH_{m}^{s} \bR_{X_{m} X_{m}} {\bH_{m}^{s}}^T +
\bR_{\zeta_{m}^{s}\zeta_{m}^{s}}) \nonumber\\
&\geq \min_{s,m}\lambda_{\min}(\bR_{\zeta_{m}^{s}\zeta_{m}^{s}}).
\end{align}
The last inequality holds because (i) for any matrices $\bA$ and $\bC$
$\lambda_{\min}(\bA+\bC) \geq \lambda_{\min}(\bA) +
\lambda_{\min}(\bC)$ and (ii) ${\bH_{m}^{s}} \bR_{X_{m} X_{m}} {\bH_{m}^{s}}^T$
is positive semidefinite.

Similarly, since $\bR_{\mathcal{Z}_{k}
\mathcal{Z}_{k}}$ is a covariance matrix and as such is positive
semidefinite we deduce:
\begin{align} \label{eqn:technical_5}
\lambda_{\min}(\bR_{\mathcal{Z}_{k} \mathcal{Z}_{k}}) &=
\lambda_{\min}(\bB_{\mathcal{Z}_{k} \mathcal{Z}_{k}} +
(\bR_{\mathcal{Z}_{k} \mathcal{Z}_{k}}-\bB_{\mathcal{Z}_{k}
\mathcal{Z}_{k}})) \\
&\geq \max[0, \lambda_{\min}(\bB_{\mathcal{Z}_{k} \mathcal{Z}_{k}})
+ \lambda_{\min}(\bR_{\mathcal{Z}_{k}
\mathcal{Z}_{k}}-\bB_{\mathcal{Z}_{k} \mathcal{Z}_{k}})] \\
&\geq \min_{s,m}\lambda_{\min}(\bR_{\zeta_{m}^{s}\zeta_{m}^{s}}) -
\rho(\bB_{\mathcal{Z}_{k} \mathcal{Z}_{k}}-\bR_{\mathcal{Z}_{k}
\mathcal{Z}_{k}})
\end{align}
The last line is valid provided $\min_{s,m}\lambda_{\min}(\bR_{\zeta_{m}^{s}\zeta_{m}^{s}}) >
\rho(\bB_{\mathcal{Z}_{k} \mathcal{Z}_{k}}-\bR_{\mathcal{Z}_{k}
\mathcal{Z}_{k}})$, which holds for sufficiently large
$\lambda_{\min}$ due to the finite bound derived for the spectral
radius in \eqref{eqn:technical_4}. 

We can now derive the following bound on the expression of interest in
the theorem, employing the relationship $\tr(\cdot) = \sum_i \lambda_i(\cdot)$:
\begin{align} \label{eqn:technical_0}
| \tr \bR_{X_{k} \mathcal{Z}_{k}}\bR_{\mathcal{Z}_{k}
\mathcal{Z}_{k}}^{-1}\bR_{\mathcal{Z}_{k} X_{k}} &-  \tr \bR_{X_{k}
\mathcal{Z}_{k}}\bB_{\mathcal{Z}_{k}
\mathcal{Z}_{k}}^{-1}\bR_{\mathcal{Z}_{k} X_{k}} |\\
 &= | \tr
\bR_{X_{k} \mathcal{Z}_{k}}(\bR_{\mathcal{Z}_{k}
\mathcal{Z}_{k}}^{-1}-\bB_{\mathcal{Z}_{k}
\mathcal{Z}_{k}}^{-1})\bR_{\mathcal{Z}_{k} X_{k}}| \\
&\leq K \ell \rho(\bR_{X_{k} \mathcal{Z}_{k}}(\bR_{\mathcal{Z}_{k}
\mathcal{Z}_{k}}^{-1}-\bB_{\mathcal{Z}_{k}
\mathcal{Z}_{k}}^{-1})\bR_{\mathcal{Z}_{k} X_{k}}),\\
&\leq K \ell \rho(\bR_{\mathcal{Z}_{k} \mathcal{Z}_{k}}^{-1}-\bB_{\mathcal{Z}_{k}
\mathcal{Z}_{k}}^{-1})\rho(\bR_{X_{k}
\mathcal{Z}_{k}}\bR_{\mathcal{Z}_{k} X_{k}}), \label{eqn:th1_a}\\
&\leq K \ell \lambda_{\min}^{-1}(\bR_{\mathcal{Z}_{k} \mathcal{Z}_{k}})
\lambda_{\min}^{-1}(\bB_{\mathcal{Z}_{k} \mathcal{Z}_{k}}) \rho(
\bB_{\mathcal{Z}_{k} \mathcal{Z}_{k}} - \bR_{\mathcal{Z}_{k}
\mathcal{Z}_{k}}) \rho(\bR_{X_{k}
\mathcal{Z}_{k}}\bR_{\mathcal{Z}_{k} X_{k}}),\label{eqn:th1_b}\\
&\leq \frac{K\ell\rho(\bR_{X_{k} \mathcal{Z}_{k}}\bR_{\mathcal{Z}_{k} X_{k}})
\rho( \bB_{\mathcal{Z}_{k} \mathcal{Z}_{k}} - \bR_{\mathcal{Z}_{k}
\mathcal{Z}_{k}})}{(\min_{s,m}\lambda_{\min}(\bR_{\zeta_{m}^{s}\zeta_{m}^{s}})
- \rho(\bB_{\mathcal{Z}_{k} \mathcal{Z}_{k}}-\bR_{\mathcal{Z}_{k}
\mathcal{Z}_{k}}))
\min_{s,m}\lambda_{\min}(\bR_{\zeta_{m}^{s}\zeta_{m}^{s}})}.
\end{align}

The finite bounds on the expressions in the numerator lead us to the conclusion that
$\min_{s,m}\lambda_{\min}(\bR_{\zeta_{m}^{s}\zeta_{m}^{s}})
\rightarrow \infty \Rightarrow | \tr \bR_{X_{k} \mathcal{Z}_{k}}\bR_{\mathcal{Z}_{k}
\mathcal{Z}_{k}}^{-1}\bR_{\mathcal{Z}_{k} X_{k}} -  \tr \bR_{X_{k}
\mathcal{Z}_{k}}\bB_{\mathcal{Z}_{k}
\mathcal{Z}_{k}}^{-1}\bR_{\mathcal{Z}_{k} X_{k}} | \rightarrow 0$, completing the proof.

\end{IEEEproof}

\bibliographystyle{IEEEtran}
\bibliography{IEEEabrv,ref}

\end{document}